
%
%

\magnification=\magstephalf
\hsize=13.0 true cm
\vsize=19 true cm
\hoffset=1.50 true cm
\voffset=2.0 true cm

\abovedisplayskip=12pt plus 3pt minus 3pt
\belowdisplayskip=12pt plus 3pt minus 3pt
\parindent=2em


\font\sixrm=cmr6
\font\eightrm=cmr8
\font\ninerm=cmr9

\font\sixi=cmmi6
\font\eighti=cmmi8
\font\ninei=cmmi9

\font\sixsy=cmsy6
\font\eightsy=cmsy8
\font\ninesy=cmsy9

\font\sixbf=cmbx6
\font\eightbf=cmbx8
\font\ninebf=cmbx9

\font\eightit=cmti8
\font\nineit=cmti9

\font\eightsl=cmsl8
\font\ninesl=cmsl9

\font\sixss=cmss8 at 8 true pt
\font\sevenss=cmss9 at 9 true pt
\font\eightss=cmss8
\font\niness=cmss9
\font\tenss=cmss10

\font\bigrm=cmr10 at 12 true pt
\font\bigbf=cmbx10 at 12 true pt

\catcode`@=11
\newfam\ssfam

\def\tenpoint{\def\rm{\fam0\tenrm}%
    \textfont0=\tenrm \scriptfont0=\sevenrm \scriptscriptfont0=\fiverm
    \textfont1=\teni  \scriptfont1=\seveni  \scriptscriptfont1=\fivei
    \textfont2=\tensy \scriptfont2=\sevensy \scriptscriptfont2=\fivesy
    \textfont3=\tenex \scriptfont3=\tenex   \scriptscriptfont3=\tenex
    \textfont\itfam=\tenit                  \def\it{\fam\itfam\tenit}%
    \textfont\slfam=\tensl                  \def\sl{\fam\slfam\tensl}%
    \textfont\bffam=\tenbf \scriptfont\bffam=\sevenbf
    \scriptscriptfont\bffam=\fivebf
                                            \def\bf{\fam\bffam\tenbf}%
    \textfont\ssfam=\tenss \scriptfont\ssfam=\sevenss
    \scriptscriptfont\ssfam=\sevenss
                                            \def\ss{\fam\ssfam\tenss}%
    \normalbaselineskip=13pt
    \setbox\strutbox=\hbox{\vrule height8.5pt depth3.5pt width0pt}%
    \let\big=\tenbig
    \normalbaselines\rm}

\def\ninepoint{\def\rm{\fam0\ninerm}%
    \textfont0=\ninerm      \scriptfont0=\sixrm
                            \scriptscriptfont0=\fiverm
    \textfont1=\ninei       \scriptfont1=\sixi
                            \scriptscriptfont1=\fivei
    \textfont2=\ninesy      \scriptfont2=\sixsy
                            \scriptscriptfont2=\fivesy
    \textfont3=\tenex       \scriptfont3=\tenex
                            \scriptscriptfont3=\tenex
    \textfont\itfam=\nineit \def\it{\fam\itfam\nineit}%
    \textfont\slfam=\ninesl \def\sl{\fam\slfam\ninesl}%
    \textfont\bffam=\ninebf \scriptfont\bffam=\sixbf
                            \scriptscriptfont\bffam=\fivebf
                            \def\bf{\fam\bffam\ninebf}%
    \textfont\ssfam=\niness \scriptfont\ssfam=\sixss
                            \scriptscriptfont\ssfam=\sixss
                            \def\ss{\fam\ssfam\niness}%
    \normalbaselineskip=12pt
    \setbox\strutbox=\hbox{\vrule height8.0pt depth3.0pt width0pt}%
    \let\big=\ninebig
    \normalbaselines\rm}

\def\eightpoint{\def\rm{\fam0\eightrm}%
    \textfont0=\eightrm      \scriptfont0=\sixrm
                             \scriptscriptfont0=\fiverm
    \textfont1=\eighti       \scriptfont1=\sixi
                             \scriptscriptfont1=\fivei
    \textfont2=\eightsy      \scriptfont2=\sixsy
                             \scriptscriptfont2=\fivesy
    \textfont3=\tenex        \scriptfont3=\tenex
                             \scriptscriptfont3=\tenex
    \textfont\itfam=\eightit \def\it{\fam\itfam\eightit}%
    \textfont\slfam=\eightsl \def\sl{\fam\slfam\eightsl}%
    \textfont\bffam=\eightbf \scriptfont\bffam=\sixbf
                             \scriptscriptfont\bffam=\fivebf
                             \def\bf{\fam\bffam\eightbf}%
    \textfont\ssfam=\eightss \scriptfont\ssfam=\sixss
                             \scriptscriptfont\ssfam=\sixss
                             \def\ss{\fam\ssfam\eightss}%
    \normalbaselineskip=10pt
    \setbox\strutbox=\hbox{\vrule height7.0pt depth2.0pt width0pt}%
    \let\big=\eightbig
    \normalbaselines\rm}

\def\tenbig#1{{\hbox{$\left#1\vbox to8.5pt{}\right.\n@space$}}}
\def\ninebig#1{{\hbox{$\textfont0=\tenrm\textfont2=\tensy
                       \left#1\vbox to7.25pt{}\right.\n@space$}}}
\def\eightbig#1{{\hbox{$\textfont0=\ninerm\textfont2=\ninesy
                       \left#1\vbox to6.5pt{}\right.\n@space$}}}

\font\sectionfont=cmbx10
\font\subsectionfont=cmti10

\def\figurecaptionfont{\ninepoint}
\def\tablecaptionfont{\ninepoint}
\def\footnotefont{\eightpoint}


\newcount\equationno
\newcount\bibitemno
\newcount\figureno
\newcount\tableno

\equationno=0
\bibitemno=0
\figureno=0
\tableno=0
\advance\pageno by -1


\footline={\ifnum\pageno=0{\hfil}\else
{\hss\rm\the\pageno\hss}\fi}


\def\section #1. #2 \par
{\vskip0pt plus .20\vsize\penalty-150 \vskip0pt plus-.20\vsize
\vskip 1.6 true cm plus 0.2 true cm minus 0.2 true cm
\global\def\equationlabel{#1}
\global\equationno=0
\centerline{\sectionfont #1. #2}\par
\immediate\write\terminal{Section #1. #2}
\vskip 0.7 true cm plus 0.1 true cm minus 0.1 true cm}


\def\subsection #1 \par
{\vskip0pt plus .15\vsize\penalty-50 \vskip0pt plus-.15\vsize
\vskip2.5ex plus 0.1ex minus 0.1ex
\leftline{\subsectionfont #1}\par
\immediate\write\terminal{Subsection #1}
\vskip1.0ex plus 0.1ex minus 0.1ex
\noindent}


\def\appendix #1 \par
{\vskip0pt plus .20\vsize\penalty-150 \vskip0pt plus-.20\vsize
\vskip 1.6 true cm plus 0.2 true cm minus 0.2 true cm
\global\def\equationlabel{\hbox{\rm#1}}
\global\equationno=0
\centerline{\sectionfont Appendix #1}\par
\immediate\write\terminal{Appendix #1}
\vskip 0.7 true cm plus 0.1 true cm minus 0.1 true cm}


\def\enum{\global\advance\equationno by 1
(\equationlabel.\the\equationno)}


\def\ifundefined#1{\expandafter\ifx\csname#1\endcsname\relax}

\def\ref#1{\ifundefined{#1}?\immediate\write\terminal{unknown reference
on page \the\pageno}\else\csname#1\endcsname\fi}

\newwrite\terminal
\newwrite\bibitemlist

\def\bibitem#1#2\par{\global\advance\bibitemno by 1
\immediate\write\bibitemlist{\string\def
\expandafter\string\csname#1\endcsname
{\the\bibitemno}}
\item{[\the\bibitemno]}#2\par}

\def\beginbibliography{
\vskip0pt plus .20\vsize\penalty-150 \vskip0pt plus-.20\vsize
\vskip 1.6 true cm plus 0.2 true cm minus 0.2 true cm
\centerline{\sectionfont References}\par
\immediate\write\terminal{References}
\immediate\openout\bibitemlist=biblist
\frenchspacing
\vskip 0.7 true cm plus 0.1 true cm minus 0.1 true cm}

\def\endbibliography{
\immediate\closeout\bibitemlist
\nonfrenchspacing}


\def\figurecaption#1{\global\advance\figureno by 1
\narrower\figurecaptionfont
Fig.~\the\figureno. #1}

\def\tablecaption#1{\global\advance\tableno by 1
\vbox to 0.5 true cm { }
\centerline{\tablecaptionfont%
Table~\the\tableno. #1}
\vskip-0.4 true cm}

\tenpoint


\def\blackboardrrm{\mathchoice
{\rm I\kern-0.21 em{R}}{\rm I\kern-0.21 em{R}}
{\rm I\kern-0.19 em{R}}{\rm I\kern-0.19 em{R}}}

\def\blackboardzrm{\mathchoice
{\rm Z\kern-0.32 em{Z}}{\rm Z\kern-0.32 em{Z}}
{\rm Z\kern-0.28 em{Z}}{\rm Z\kern-0.28 em{Z}}}

\def\blackboardh{\mathchoice
{\ss I\kern-0.14 em{H}}{\ss I\kern-0.14 em{H}}
{\ss I\kern-0.11 em{H}}{\ss I\kern-0.11 em{H}}}

\def\blackboardp{\mathchoice
{\ss I\kern-0.14 em{P}}{\ss I\kern-0.14 em{P}}
{\ss I\kern-0.11 em{P}}{\ss I\kern-0.11 em{P}}}

\def\blackboardt{\mathchoice
{\ss T\kern-0.52 em{T}}{\ss T\kern-0.52 em{T}}
{\ss T\kern-0.40 em{T}}{\ss T\kern-0.40 em{T}}}

\def\thicktablerule{\hrule height1pt}
\def\thintablerule{\hrule height0.4pt}




\def\rmd{{\rm d}}
\def\rmD{{\rm D}}
\def\rme{{\rm e}}
\def\rmO{{\rm O}}


\def\rz{\blackboardrrm}
\def\gz{\blackboardzrm}

\def\Re{{\rm Re}\,}


\def\proof{\noindent{\sl Proof:}\kern0.6em}
\def\endproof{\hskip0.6em plus0.1em minus0.1em
\setbox0=\null\ht0=6.4pt\dp0=1pt\wd0=6.3pt
\vbox{\hrule height0.8pt
\hbox{\vrule width0.8pt\box0\vrule width0.8pt}
\hrule height0.8pt}}
\def\frac#1#2{\hbox{$#1\over#2$}}
\def\dual{\mathstrut^*\kern-0.1em}



\def\euler{\gamma_{\rm E}}


\def\gms{g_{\ms}}
\def\gbar{\bar{g}}
\def\gbarms{\gbar_{\ms}}

\def\ms{{\rm MS}}
\def\msbar{{\rm \overline{MS\kern-0.14em}\kern0.14em}}
\def\lat{{\rm lat}}
\def\glat{g_{\lat}}


\def\SU{{\rm SU}(N)}
\def\SUtwo{{\rm SU}(2)}
\def\SUthree{{\rm SU}(3)}
\def\su{{\rm su}(N)}

\def\pauli#1{\tau^{#1}}
\def\Ad{{\rm Ad}\,}


\def\schrodinger{{\cal Z}}
\def\effaction{\Gamma}
\def\chernsimons{S_{\rm CS}}


\def\bvalue{C}
\def\bfield{B}
\def\bfieldtensor{G}
\def\bfunc{b}
\def\bfieldparm{\eta}
\def\qfield{q}
\def\bvaluelat{W}
\def\bfieldlat{V}
\def\bfunclat{v}


\def\deltazero{\Delta_0}
\def\deltazerohat{\hat{\Delta}_0}
\def\deltaone{\Delta_1}
\def\deltaonehat{\hat{\Delta}_1}
\def\deltaoneprime{\Delta\kern-1.0pt
    \smash{\raise 4.5pt\hbox{$\scriptstyle\prime$}}
    \kern-1.5pt_{1}}
\def\FP{{\rm FP}}
\def\deltaSone{\Delta^{{\rm S}^1}}


\def\seeleycoeff#1#2{\alpha_{#1}({#2})}
\def\zetafunc#1#2{\zeta(#1|#2)}
\def\zetaprime#1{\zeta'(0|#1)}


\def\ham{\blackboardh}
\def\projector{\blackboardp}
\def\trans{\blackboardt}
\def\Tr{{\rm Tr}}
\def\tr{{\rm tr}}


\def\delstar#1{\Delta\kern-1.0pt\smash{\raise 4.5pt\hbox{$\ast$}}
               \kern-4.0pt_{#1}}

\def\nabstar#1{\nabla\kern-0.5pt\smash{\raise 4.5pt\hbox{$\ast$}}
               \kern-4.5pt_{#1}}
\def\cdev#1{D\kern-0.2pt\smash{\raise 4.2pt
            \hbox{$\scriptstyle\phantom{\ast}$}}
            \kern-4.8pt_{#1}}
\def\cdevstar#1{D\kern-0.2pt\smash{\raise 4.2pt
                \hbox{$\scriptstyle\ast$}}
                \kern-4.8pt_{#1}}
\def\plaq{P}
\def\dplaq{d^*\!P}


\def\angle{\phi^{\vphantom{\prime}}}
\def\angleprime{\phi'}
\def\anglevector{\Phi^{\vphantom{\prime}}}
\def\anglevectorprime{\Phi'}
\def\temptrans{\Theta}
\def\setS{{\cal S}}
\def\perm{\sigma}


\def\gaugegroup{{\cal G}}
\def\largegaugegroup{\hat{\gaugegroup}}
\def\cartangroup{{\rm C}_N}
\def\gaugealgebra{{\cal L}}
\def\fluctuationspace{{\cal H}}
\def\ga#1{\gaugealgebra_{#1}}
\def\fs#1{\fluctuationspace_{#1}}
\def\gap#1{\ga{#1}({\bf p})}
\def\fsp#1{\fs{#1}({\bf p})}
\def\gaugefunction{F}


\def\amat{{\cal A}}
\def\bmat{{\cal B}}
\def\cmat{{\cal C}}
\def\solmat{M}
\def\phat{\hat{p}}
\def\cgh{\cos{\gamma\over2}}
\def\funcspace{{\cal F}}
\def\dim{d_{\funcspace}}

%
\vskip 4 true cm
\centerline
{\bigbf The Schr\"odinger Functional --- a Renormalizable Probe}
\centerline
{\bigbf for Non-Abelian Gauge Theories}
\vskip 2 true cm
\centerline{\bigrm Martin L\"{u}scher and Rajamani Narayanan}
\vskip 2ex
\centerline{Deutsches Elektronen-Synchrotron DESY}
\centerline{Notkestrasse 85, D-2000 Hamburg 52, Germany}
\vskip 1 true cm
\centerline{\bigrm Peter Weisz}
\vskip 2ex
\centerline{Max Planck Institut f\"ur Physik}
\centerline{F\"ohringerring 6, D-8000 M\"unchen 40, Germany}
\vskip 1 true cm
\centerline{\bigrm Ulli Wolff}
\vskip 2ex
\centerline{CERN, Theory Division, CH-1211 Gen\`eve 23, Switzerland}
\vskip 2.0 true cm
\centerline{\bf Abstract}
\vskip 1.5ex
Following Symanzik we argue that the
Schr\"odinger functional in
lattice gauge theories without matter fields
has a well-defined continuum limit.
Due to gauge invariance no extra counter terms are required.
The Schr\"odinger functional is, moreover, accessible to
numerical simulations.
It may hence be used to study the scaling properties of the theory
and in particular the evolution of
the renormalized gauge coupling from low to high energies.
A concrete proposition along this line is made and the necessary
perturbative analysis of the Schr\"odinger functional is
carried through to 1-loop order.
\vfill
\eject

\section 1. Introduction

The Schr\"odinger functional is the propagation kernel
for going from some field configuration at time $x^0=0$
to some other configuration at $x^0=T$. In euclidean space-time
it can be written as a functional integral over
all fields with the specified initial and final values.
The renormalization of the Schr\"odinger functional has
been discussed by Symanzik [\ref{SymSchrodinger}] some
time ago in the course of his proof of the existence of the
Schr\"odinger picture in renormalizable quantum field theories
(for an introduction to this paper see ref.[\ref{LueSchrodinger}]).
His result is that the Schr\"odinger functional can be renormalized
by adding the usual counterterms to the action plus possibly a set of
further terms that are integrals of local
polynomials in the field and its derivatives over
the $x^0=0$ and $x^0=T$ hyper-planes.
In the case of the $\phi^4$ theory, for example, two
polynomials, $\phi^2$ and $\phi\partial_0\phi$, are needed.
Pure gauge theories are simpler in this respect, because
no extra counterterms are required here, as we shall argue in
sect.~2. If we choose a lattice to regularize the theory, this
statement simply means that the Schr\"odinger
functional converges in the
continuum limit, provided, of course, the bare coupling is scaled in
the usual way.

Our motivation to consider
the Schr\"odinger functional is that
we would like to apply the
finite size scaling technique of
ref.[\ref{LueWeWo}] to gauge theories.
The ultimate goal of this programme
is to compute the running coupling in say the
minimal subtraction ($\ms$) scheme
of dimensional regularization
at short distances given in units of the low energy scales of the
theory (the string tension, for example, or the mass of the lightest
glueball).
In other words, our aim is to connect the
non-perturbative
infrared behaviour of the theory with the high energy regime,
where the coupling is logarithmically decreasing according to
the renormalization group.

To explain in which way the Schr\"odinger functional
enters this calculation, we need to recall
the basic strategy of ref.[\ref{LueWeWo}].
One begins by putting the theory in a finite spatial
volume with linear extension $L$ and periodic boundary conditions
in all directions. Next one introduces
some renormalized coupling $\gbar^2(L)$ which does not depend
on any scale other than $L$ and which can hence be considered
a running coupling. This coupling is then computed
over a range of $L$ through numerical simulation of the
lattice theory, using a recursive procedure which allows one
to go from large values of $L$ (where contact is made
with the non-perturbative scales) to the perturbative
small $L$ domain. At these distances
$\gbar^2(L)$ can be analytically related
to other more commonly used couplings such as
the coupling in the $\ms$ scheme of dimensional
regularization.
Note that
the latter are usually defined in infinite volume.
All reference to a finite volume thus disappears from the
final result.

The precise definition of the running coupling $\gbar^2(L)$
is not of principal importance.
The coupling should however be accurately computable
through numerical simulation and its scaling properties
should not be strongly influenced by the presence of
a non-zero lattice spacing.
It is our experience that these requirements are not easy to
fulfil.
In particular, extracting a running coupling from
correlation functions of Wilson loops is difficult,
because large loops give a poor signal while small loops
are affected by lattice artefacts.
As we shall explain in sect.~2,
running couplings can be straightforwardly defined through
the Schr\"odinger functional. With carefully chosen
boundary values for the gauge field, these couplings
have, moreover, the desired technical properties which make our
finite size scaling study feasible.

The idea to probe gauge theories through external fields
is not new of course. In particular,
the background field method
[\ref{DeWittBackground}--\ref{Abbott}]
has long proved to
be an efficient tool for perturbative
calculations, in a wide range of theories and contexts
(see e.g.~refs.[\ref{DashenGross}--\ref{VenBackground}]).
Our work can also be understood
as a continuation
of ref.[\ref{WoGbar}].
Most of the basic concepts
that we now exploit already
appear there, in a language oriented towards statistical
mechanics.

In the present paper
the Schr\"odinger functional in Yang-Mills theories
is discussed in detail.
Numerical results on
the scaling behaviour of the
couplings derived from it will be published elsewhere.
After going through the formal definition of the functional in
sect.~2, we verify explicitly, using dimensional regularization,
that indeed there are no additional
counterterms at one-loop order
of perturbation theory (sect.~3).
We then discuss
the definition and cutoff dependence
of the Schr\"odinger functional
on the lattice (sect.~4).
In our numerical work, we take
constant Abelian fields for the
boundary values of the gauge field at $x^0=0$ and $x^0=T$,
and so, in sects.~5 and 6,
we set up the perturbation expansion
for this case.
In particular, the stability of the gauge field configuration
around which one expands
must be established and
the gauge fixing must be done with care
to obtain all boundary terms correctly.
As an application we calculate
our running coupling to one-loop order,
for the theory with gauge group $\SUtwo$ (sect.~7).


\section 2. A first look at the Schr\"odinger functional

The aim in this section is to give an introduction to the
Schr\"odinger functional in Yang-Mills theories
and to establish the basic notation.
We shall not at this point worry about the mathematical
status of the quantities considered. Later on when the
theory is regularized either dimensionally or by passing to
the lattice formulation, we shall be able to
deal with the Schr\"odinger functional on
a more rigorous level.

For definiteness we assume that the gauge group is $\SU$.
Our conventions regarding indices, group generators etc.~are
listed in appendix A.

\subsection 2.1 Formal definition of the Schr\"odinger functional

Our starting point is the
Hamiltonian formulation of
the theory in the temporal gauge.
In this framework the theory is specified at a fixed time, say
$x^0=0$, by assuming canonical commutation relations among
the basic field variables and by giving the Hamilton operator.

As explained in the introduction,
the Schr\"odinger functional will be used to study the scaling
properties of the theory
in finite volume, and so we take space to
be an $L\times L\times L$ box with
periodic boundary conditions.
$\SU$ gauge fields are accordingly
represented by periodic
vector potentials $A_k({\bf x})$ on $\rz^3$ with values in the
Lie algebra of $\SU$
\footnote{$\dag$}{\footnotefont
We do not consider the possibility of twisted periodic boundary
conditions in this paper.}.
To preserve periodicity
under gauge transformations
$$
  A_k({\bf x})\to
  A_k^{\Lambda}({\bf x})=
  \Lambda({\bf x})A_k({\bf x})\Lambda({\bf x})^{-1}+
  \Lambda({\bf x})\partial_k\Lambda({\bf x})^{-1},
  \eqno\enum
$$
only periodic gauge functions
$\Lambda({\bf x})$ will be admitted.
$\Lambda$ can thus
be regarded as a mapping from a 3-dimensional
torus to $\SU$. Continuous functions of this kind are
topologically non-trivial in general.
More precisely,
they fall into disconnected topological classes
labelled by an integer winding number
$$
  n
  ={1\over24\pi^2}\int_0^L\rmd^3x\,\epsilon_{klj}\,
  \tr\left\{\left(\Lambda\partial_k\Lambda^{-1}\right)
            \left(\Lambda\partial_l\Lambda^{-1}\right)
            \left(\Lambda\partial_j\Lambda^{-1}\right)\right\}.
  \eqno\enum
$$
Gauge transformations with non-zero winding number are
explicitly allowed
and will play an important r\^ole later on.

In the Schr\"odinger representation the quantum mechanical
states of the theory are
wave functionals $\psi[A]$, where $A$ runs
over all gauge fields as described above.
A scalar product is formally given by
$$
  \langle\psi|\chi\rangle=\int\rmD[A]\,\psi[A]^*\chi[A],
  \qquad\rmD[A]=\prod_{{\bf x},k,a}\rmd A_k^a({\bf x}).
  \eqno\enum
$$
Only gauge invariant
states $\psi[A]$,
i.e.~those satisfying
$$
  \psi[A^{\Lambda}]=\psi[A]
  \eqno\enum
$$
for all gauge transformations $\Lambda$, are physical.
In particular, we choose the vacuum angle $\theta$
[\ref{JaRe},\ref{CaDaGr}] to vanish.
Any given wave functional $\psi[A]$ can be projected on
the physical subspace through
$$
  \psi[A]\to
  \projector\psi[A]=\int\rmD[\Lambda]\,\psi[A^\Lambda],
  \qquad\rmD[\Lambda]=\prod_{\bf x}\rmd\Lambda({\bf x}),
  \eqno\enum
$$
where $\rmd U$, $U\in\SU$,
denotes the normalized invariant measure on $\SU$.

The gauge field $A_k^a({\bf x})$ can in the obvious way
be interpreted as an operator field
acting on wave functionals
$\psi[A]$. The canonically conjugate field is the
colour electric field
$$
  F_{0k}^a({\bf x})
  ={1\over i}{\delta\over\delta A_k^a({\bf x})}.
  \eqno\enum
$$
The magnetic components of the colour field tensor are
$$
  F_{kl}^a({\bf x})
  =\partial_kA_l^a({\bf x})-\partial_lA_k^a({\bf x})
   +f^{abc}A_k^b({\bf x})A_l^c({\bf x}),
  \eqno\enum
$$
and the Hamilton operator $\ham$ is given by
$$
  \ham
  =\int_0^L\rmd^3x\,\left\{
  {g_0^2\over2}F_{0k}^a({\bf x})F_{0k}^a({\bf x})+
  {1\over4g_0^2}F_{kl}^a({\bf x})F_{kl}^a({\bf x})\right\}
  \eqno\enum
$$
with $g_0$ being the (bare) gauge coupling.

For any smooth classical gauge field
$\bvalue_k({\bf x})$, a state
$|\bvalue\rangle$ may be introduced such that
$$
  \langle \bvalue|\psi\rangle=\psi[\bvalue]
  \eqno\enum
$$
for all wave functionals $\psi[A]$.
$|\bvalue\rangle$ is not gauge invariant,
of course,
but it may be made so by applying the projector $\projector$.
The (euclidean) Schr\"odinger functional
$\schrodinger[\bvalue',\bvalue]$
is now defined by
$$
  \schrodinger[\bvalue',\bvalue]=
  \langle \bvalue'|\rme^{-\ham T}\projector|\bvalue\rangle.
  \eqno\enum
$$
We shall always assume $T>0$ and do not
explicitly indicate the dependence of
the Schr\"odinger functional on this parameter.
If we insert an orthonormal
basis $|\psi_n\rangle$, $n=0,1,2,\ldots$, of
gauge invariant energy eigenstates, the spectral representation
$$
  \schrodinger[\bvalue',\bvalue]=
  \sum_{n=0}^{\infty} \rme^{-E_nT}
  \psi_n[\bvalue'] \psi_n[\bvalue]^*
  \eqno\enum
$$
is obtained, where $E_n$ are the energy eigenvalues
(the spectrum is discrete in finite volume).
Since only physical intermediate states contribute,
it is evident that
$\schrodinger[\bvalue',\bvalue]$
is invariant under arbitrary
gauge transformations of the boundary
fields $\bvalue$ and $\bvalue'$.

\subsection 2.2 Functional integral representation

The matrix elements of the euclidean time evolution operator
$\rme^{-\ham T}$ between gauge invariant states can be expressed
through a functional integral over all
gauge field configurations
$A_{\mu}(x)$ in four dimensions with $0\leq x^0\leq T$
and periodic boundary conditions in the spatial directions.
For the matrix element (2.10), the appropriate
initial and final values of the gauge field are
$$
  A_k(x)=\cases{
  \bvalue_k^{\Lambda}({\bf x}) & at $x^0=0$,\cr
  \noalign{\vskip1ex}
  \bvalue'_k({\bf x})          & at $x^0=T$,\cr}
  \eqno\enum
$$
and an integration over all gauge transformations
$\Lambda$ is required to account for the projector
$\projector$ [cf.~eq.(2.5)].
The functional integral representation of the
Schr\"odinger functional thus reads
$$
  \schrodinger[\bvalue',\bvalue]=
  \int\rmD[\Lambda]\int\rmD[A]\,
  \rme^{-S[A]},
  \eqno\enum
$$
where the measure $\rmD[A]$ now stands for an integration
over all components of the euclidean field.
The euclidean action is given by
$$
  S[A]=-{1\over2g_0^2}\int\rmd^4x\,
  \tr\left\{F_{\mu\nu}F_{\mu\nu}\right\},
  \eqno\enum
$$
with
$$
  F_{\mu\nu}=\partial_{\mu}A_{\nu}-\partial_{\nu}A_{\mu}
  +[A_{\mu},A_{\nu}].
  \eqno\enum
$$
Finally, an overall normalization factor independent
of the boundary values $\bvalue$ and $\bvalue'$ has been dropped
in eq.(2.13). In the following
we do not keep track of such factors,
because we are ultimately only interested in the
field dependence
of the Schr\"odinger functional.

The reader may wonder at this point how it comes that
the time component of the gauge field appears in the functional
integral, while in the Hamiltonian formulation it did not.
The reason is that the integral (2.13) and the boundary conditions
(2.12) are invariant
under the gauge transformation
$$
  \eqalignno{
  A_{\mu}(x)&\to
  \Omega(x)A_{\mu}(x)\Omega(x)^{-1}+
  \Omega(x)\partial_{\mu}\Omega(x)^{-1},&\enum\cr
  \noalign{\vskip1ex}
  \Lambda({\bf x})&\to
  \Omega(x)|_{x^0=0}\Lambda({\bf x}),&\enum\cr}
$$
provided the gauge function $\Omega(x)$ satisfies
$$
  \Omega(x)=1\quad\hbox{at}\quad x^0=T.
  \eqno\enum
$$
An admissible gauge fixing condition for this symmetry is the
temporal gauge $A_0=0$.
The associated Faddeev-Popov determinant
is field independent and may be included
in the overall normalization of the integral.
The time component
of the gauge field can thus be eliminated
and it is now straightforward to
make contact with the Hamiltonian
expression for the Schr\"odinger functional.

For the perturbative calculations to be discussed later on,
other gauge choices will be more convenient so that in the
following
we keep the time component of the gauge field as an integration
variable.

We may however
use the gauge invariance of the functional integral to
reduce the integration over the gauge transformation
$\Lambda$ to a sum over topological classes.
To this end, first note that after
the inner integral in eq.(2.13) has been performed, one
is left with some function of
$\Lambda$ to be integrated over.
This function depends only on the
winding number of $\Lambda$,
because any two gauge functions in the same topological
class may be connected through a gauge transformation
$\Omega$ as described above [eqs.(2.16--18)].
So if we choose, for each integer $n$,
some fixed representative
$\Lambda_n({\bf x})$ in the class of gauge functions with
winding number $n$,
the Schr\"odinger functional becomes
$$
  \schrodinger[\bvalue',\bvalue]=
  \sum_{n=-\infty}^{\infty}\int\rmD[A]\,
  \rme^{-S[A]},
  \eqno\enum
$$
where now we require that
$$
  A_k(x)=\cases{
  \bvalue_k^{\Lambda_n}({\bf x}) & at $x^0=0$,\cr
  \noalign{\vskip1ex}
  \bvalue'_k({\bf x})            & at $x^0=T$.\cr}
  \eqno\enum
$$
It is convenient to set $\Lambda_0=1$, but an explicit
choice of the non-trivial
$\Lambda_n$'s would not be useful in what follows.
That a sum over topological classes is needed on top of
the functional integration is perhaps not too surprising,
if we recall that this is the normal situation
in Yang-Mills theories on compact
manifolds without boundary.

\subsection 2.3 Instanton bound

It is well-known [\ref{BePo}] that
the gauge field action
is bounded by
$$
  S[A]\geq{8\pi^2\over g_0^2}\left|Q[A]\right|,
  \eqno\enum
$$
where
$$
  Q[A]=-{1\over16\pi^2}\int\rmd^4x\,
  \tr\left\{F_{\mu\nu}\dual F_{\mu\nu}\right\}
  \eqno\enum
$$
denotes the topological charge of $A$ and
$$
  \dual F_{\mu\nu}=
  \frac{1}{2}\epsilon_{\mu\nu\rho\sigma}
  F_{\rho\sigma}
  \eqno\enum
$$
the dual of the field tensor (2.15).
$Q[A]$ may be expressed through
the boundary values $\bvalue$ and $\bvalue'$
and the winding number $n$.
To see this note that
$$
  \frac{1}{2}
  \tr\left\{F_{\mu\nu}\dual F_{\mu\nu}\right\}
  =\epsilon_{\mu\nu\rho\sigma}\partial_{\mu}
  \tr\left\{A_{\nu}\partial_{\rho}A_{\sigma}+
  \frac{2}{3}A_{\nu}A_{\rho}A_{\sigma}\right\}.
  \eqno\enum
$$
Taking the boundary conditions (2.20) into account and integrating
by parts then yields (after some algebra)
$$
  Q[A]=\chernsimons[\bvalue']-\chernsimons[\bvalue]-n,
  \eqno\enum
$$
where
$$
  \chernsimons[\bvalue]=
  -{1\over8\pi^2}\int\rmd^3x\,\epsilon_{klj}\,
  \tr\left\{\bvalue_k\partial_l\bvalue_j+
  \frac{2}{3}\bvalue_k\bvalue_l\bvalue_j\right\}
  \eqno\enum
$$
is the Chern-Simons action of the boundary field $\bvalue$.

\subsection 2.4 Induced background field

For small couplings $g_0$, the functional integral
(2.19) is dominated by the fields around the absolute
minima of the action.
{}From the instanton bound it follows that
we only need to inspect a small number of
topological sectors
to find the fields with least action (and the
specified boundary values).
In general there are
several gauge inequivalent
minimal action configurations and these can even occur in
different winding number sectors.

For simplicity we shall from now on
restrict attention to boundary values
$\bvalue$ and $\bvalue'$, where the
absolute minimum is attained in the $n=0$ sector
and where, furthermore, the
minimal action configuration $\bfield_{\mu}(x)$
is unique
up to gauge transformations.
This is the typical situation
if $\bvalue$ and $\bvalue'$ are small.
In the following
$\bfield$ will be referred to as the induced background field.

For given boundary values $\bvalue$ and $\bvalue'$,
it is usually impossible to obtain the induced background field
in closed analytical form.
We may, however, invert the procedure
and start from any
known solution $B$ of the field equations.
If we {\it define} $\bvalue$ and $\bvalue'$ through
$$
  \bvalue_k({\bf x})=\left.B_k(x)\right|_{x^0=0},
  \qquad
  \bvalue'_k({\bf x})=\left.B_k(x)\right|_{x^0=T},
  \eqno\enum
$$
the boundary conditions
will be trivially satisfied.
We are then left
with the problem to prove that the chosen field
$B$ is the unique minimal action configuration with these
boundary values.

Little work is needed to show this,
if the field tensor
$$
  \bfieldtensor_{\mu\nu}=\partial_{\mu}\bfield_{\nu}
                        -\partial_{\mu}\bfield_{\nu}
                        +[\bfield_{\mu},\bfield_{\nu}]
  \eqno\enum
$$
is self-dual,
$$
  \bfieldtensor_{\mu\nu}=\dual\bfieldtensor_{\mu\nu},
  \eqno\enum
$$
and if, furthermore, the bound
$$
  \chernsimons[C']-\chernsimons[C]<\frac{1}{2}
  \eqno\enum
$$
holds.
The point is that the instanton bound (2.21) is
saturated in this case.
All fields $A$ in the $n=0$ sector hence have an action
greater or equal to
$$
  S[B]={8\pi^2\over g_0^2}\left\{
  \chernsimons[\bvalue']-\chernsimons[\bvalue]\right\}.
  \eqno\enum
$$
And the same is true for all other sectors, as one may
quickly show by combining
eqs.(2.21),(2.25),(2.31) with the bound (2.30).

We have thus proved that under the conditions stated
above, $B$ is an absolute minimum of the
action. It is in fact the only one, up to gauge transformations,
because any other minimizing configuration would have to be
self-dual and satisfy the same boundary conditions.
Since the self-duality equation is first order in the time
derivatives, the solution in the temporal gauge is uniquely
determined by the initial values at $x^0=0$.
In a general gauge, this means that any two solutions
are related by a gauge transformation.

A simple example of a self-dual field is obtained by
making the ansatz
$$
  \bfield_0(x)=0,\qquad \bfield_k(x)=
  \bfunc(x^0)I_k.
  \eqno\enum
$$
Here, $\bfunc$ is a real function and the group generators $I_k$
are chosen such that they
form an irreducible representation of the angular momentum
algebra
$$
  \left[I_k,I_l\right]=\epsilon_{klj}I_j.
  \eqno\enum
$$
Up to unitary transformations, there exists only one
irreducible representation of this algebra
with dimension $N$. In particular, the square of the angular
momentum is given by
$$
  I_kI_k=-\frac{1}{4}(N^2-1).
  \eqno\enum
$$
Note that
the action of the field (2.32) is finite
if $\bfunc(x^0)$ is smooth and if $L<\infty$.

With the ansatz (2.32) the self-duality equation (2.29)
reduces to
$$
  \partial_0\bfunc=\bfunc^2,
  \eqno\enum
$$
so that
$$
  \bfunc(x^0)=(\tau-x^0)^{-1}.
  \eqno\enum
$$
To ensure the regularity of the solution in the interval
$0\leq x^0\leq T$, the integration constant $\tau$ must either be negative
or greater than $T$.
We leave it to the reader to work out the condition (2.30),
but it is obviously satisfied for
sufficiently large values of
$\tau$.
We have thus found a one-parameter family of globally stable
background fields.

\subsection 2.5 Renormalization of the Schr\"odinger functional

In the weak coupling domain, the Schr\"odinger functional
can be computed
by performing a saddle point expansion of the functional integral (2.19)
about the induced background field $\bfield$.
For the effective action
$$
  \effaction[\bfield]=
  -\ln \schrodinger\left[\bvalue',\bvalue\right],
  \eqno\enum
$$
an asymptotic series of the form
$$
  \effaction[\bfield]=
  g_0^{-2}\effaction_0[\bfield]+
  \effaction_1[\bfield]+g_0^2\effaction_2[\bfield]
  +\ldots
  \eqno\enum
$$
is thus obtained
\footnote{$\dag$}{\footnotefont
As already mentioned earlier,
we are only interested in the field
dependence of the Schr\"odinger functional. Any additive
contributions to the effective action independent of $B$
are hence dropped without further notice.}.
The leading term is given by
$$
  \effaction_0[\bfield]=g_0^2S[\bfield],
  \eqno\enum
$$
while the higher order contributions are sums of vacuum bubble
Feynman diagrams with an increasing number of loops.
The Feynman rules involve
vertices and propagators that depend on the background field.
To make the diagrams well-defined
an ultra-violet regularization will be needed at this point.
We shall either use dimensional regularization or introduce
a space-time lattice.
In both cases the gauge invariance of the theory is preserved,
which is crucial for the renormalization to work out in
the way described below.

Initially the saddle point expansion of the
Schr\"odinger functional is performed in the regularized bare
theory. That is, we integrate over the bare gauge field $A$,
use the bare action to generate the Feynman rules and
impose the boundary conditions (2.20) on the bare field.
Since the regularization respects the gauge symmetry,
the effective action $\effaction[\bfield]$
does not depend on
the gauge fixing condition employed.
Moreover, it is
invariant under gauge transformations
$$
  \bfield_{\mu}(x)\to
  \bfield_{\mu}^{\Omega}(x)=
  \Omega(x)\bfield_{\mu}(x)\Omega(x)^{-1}+
  \Omega(x)\partial_{\mu}\Omega(x)^{-1}
  \eqno\enum
$$
of the background field for arbitrary (periodic) gauge functions
$\Omega$.

Ultra-violet divergences now appear in each order of the
bare coupling $g_0$ when we try to remove the regularization.
We certainly expect that
some of the divergent terms are cancelled
by the usual renormalization of the coupling constant.
As we shall show in detail in the following sections,
the one-loop effective action actually becomes
finite after the coupling is renormalized.
In other words, what is suggested is
that the (bare) Schr\"odinger functional does not
need any renormalization besides the coupling constant
renormalization.

A rigorous proof of this conjecture to all orders of
perturbation theory is beyond the scope of this paper.
We may, however, make our proposition more plausible
by recalling
Symanzik's work on the
Schr\"odinger representation in quantum field theory
[\ref{SymSchrodinger}].
Symanzik studied the
massless $\phi^4$ theory and showed that
all divergences in the
Schr\"odinger functional can be cancelled
by renormalizing the coupling constant
and by including
the boundary counterterms
$$
  \int_{x^0=T}\rmd^3x\,
  \left\{Z_1\phi^2+Z_2\phi\partial_0\phi\right\}+
  \int_{x^0=0}\rmd^3x\,
  \left\{Z_1\phi^2-Z_2\phi\partial_0\phi\right\}
  \eqno\enum
$$
in the bare action.
The terms proportional to $\phi^2$ do not
influence the propagators and vertices
and just add to the effective action.
Because the renormalization constant $Z_1$ is linearly divergent,
they are not needed
if one employs dimensional regularization.
The other counterterms are equivalent
to a rescaling of the boundary values
of the field
[\ref{SymSchrodinger}].
They are logarithmically divergent
and must be taken into account when deriving
the renormalization group equation
for the Schr\"odinger functional.

Symanzik also expressed the expectation
that in a general renormalizable field theory,
the Schr\"odinger functional
can be made finite by renormalizing the coupling and mass parameters
and by adding
a few boundary counterterms to the action.
These are
proportional to some local composite fields of
dimension less than or equal to 3 integrated over
the $x^0=0$ and $x^0=T$ hyper-planes.

Coming back to Yang-Mills theories,
the obvious question is whether there are any
candidates for such counterterms.
Since the effective action is independent
of the gauge fixing condition employed, we
should be able to write them
in a gauge invariant form not involving the Faddeev-Popov ghosts.
We now note, however,
that any non-trivial
gauge invariant polynomial in
the gauge potential and its derivatives
has dimension greater than 3.
A counterterm proportional to
the Chern-Simons action (2.26) is also excluded, because
the latter is odd under parity.
The upshot then is that
divergent boundary terms cannot occur in this theory
and that consequently the (bare)
Schr\"odinger functional is finite after
the coupling has been renormalized.

\subsection 2.6 Running couplings in finite volume

As explained in sect.~1, the
finite size scaling study that we propose to apply to
Yang-Mills theories is based on the
idea of a renormalized coupling $\gbar^2(L)$, which runs with
the box size $L$.
Starting from
the Schr\"odinger functional, there are many ways
to introduce such a coupling.
For example, we may
choose some background field $\bfield$ which depends
on a dimensionless parameter $\bfieldparm$.
{}From the above we then infer that
$$
  \effaction'[\bfield]=
  {\partial\over\partial\bfieldparm}
  \effaction[\bfield]
  \eqno\enum
$$
is a renormalization group invariant.
A renormalized coupling may hence be defined through
$$
  \gbar^2=\effaction'_0[\bfield]\bigm/\effaction'[\bfield]
  \eqno\enum
$$
[cf.~eq.(2.38)].
Note that $\effaction'_0[\bfield]$ is just a normalizing factor,
which guarantees that
$\gbar^2$ coincides with
$g_0^2$ to leading order of perturbation theory.

In general the chosen background field depends on several
external scales
and the corresponding renormalized coupling must be regarded
as a function of all these parameters. In addition it depends
on the box size $L$ and the propagation time $T$.
To obtain a coupling which runs with $L$, we simply
scale all dimensionful
parameters with fixed proportions relative to $L$.
For example, we may set $T=L$
and take the self-dual configuration
(2.32),(2.36) with
$$
  \tau=-L/\bfieldparm
  \eqno\enum
$$
as the background field.

There are, of course, other choices of background fields
and a corresponding manifold of renormalized couplings.
In particular, a simple alternative are spatially
constant Abelian fields,
which, for reasons given later, are actually more suitable
for our numerical work.


\section 3. Perturbation Expansion

We now discuss
the saddle point expansion of the
Schr\"odinger functional around the induced background field
using dimensional regularization.
In particular, we would like to show that
to one-loop order
the renormalization of the Schr\"odinger functional
works out in the way described above.
Our calculations rely on
well-established techniques, which
have previously been
applied in the context of instantons and the semi-classical
approximation
[\ref{tHooftInstA}--\ref{MorrisInst}].
There is also a formal similarity with
the so-called background field method
[\ref{DeWittBackground}--\ref{Abbott}],
although here we do not expand in powers of
the background field.

As before we shall assume that the
absolute minimum of the action occurs in the $n=0$ sector
and that the minimal action configuration $\bfield$ is
unique up to gauge transformations.
To avoid some technicalities when fixing the gauge,
we shall in addition require that the gauge group
acts freely on the boundary values
$\bvalue$ and $\bvalue'$.
This means that the only
spatial gauge transformations
$\Lambda$ which leave $\bvalue$ or $\bvalue'$ invariant
are constant and equal to a
central element of $\SU$.
The self-dual background field considered earlier has this
property. Abelian fields however
do not and the gauge fixing procedure
must consequently
be reexamined in this case (see sect.~6).

\subsection 3.1 Dimensional regularization

The familiar techniques of dimensional regularization
apply to Feynman diagrams in the momentum space representation
[\ref{tHooftVeltman}--\ref{CiMo}].
A different approach is required
in the presence of background fields,
because the propagators
and vertices do not in general have their standard form.
One possibility is to
stay in position space and to
insert the heat kernel
representation for the propagators.
The Feynman integrals then become
well-defined analytic
functions of the space-time dimension.
In the following we shall proceed along these lines
and refer the reader to
ref.[\ref{LueDimReg}] for an introduction to the method.

Dimensional regularization starts by extending
space-time to a $D$ dimensional manifold,
where the extra $D-4$ dimensions
are here assumed to be spatial with the usual
periodic boundary conditions.
The bare action of a gauge field $A_{\mu}(x)$ on this manifold
is again given by
eq.(2.14), except that we now integrate over all $D$
coordinates of $x$ and that the Lorentz indices
run from 0 to $D-1$ (cf.~appendix A).
The boundary values $\bvalue$ and $\bvalue'$ and the background field
$\bfield$ are taken to be independent of the extra coordinates.
Their components in these directions are set to zero.
In particular,
$$
  S[\bfield]=L^{D-4}
  \left\{S[\bfield]\right\}_{D=4},
  \eqno\enum
$$
and the same volume factor also
appears in all higher order contributions
to the effective action, because the translation symmetry
in the added dimensions is not affected by the background field.

The renormalization of the gauge coupling is performed as usual.
We choose the minimal subtraction ($\ms$) scheme
[\ref{tHooftMS}]
and accordingly denote the renormalized coupling by $\gms$.
The relation between the bare and renormalized couplings
reads
$$
  g_0^2=\mu^{2\varepsilon}\gms^2
  \left\{1+\sum_{l=1}^{\infty}z_l(\varepsilon)\gms^{2l}\right\},
  \qquad D=4-2\varepsilon,
  \eqno\enum
$$
where $\mu$ is the normalization mass.
The singular coefficients
$$
  z_l(\varepsilon)=\sum_{k=1}^{l} z_{lk}\varepsilon^{-k}
  \eqno\enum
$$
coincide
with the counterterms calculated in ordinary perturbation theory.
In particular, the one-loop coefficient is given by
$$
  z_1(\varepsilon)=-{11\over3\varepsilon}{N\over16\pi^2},
  \eqno\enum
$$
and the
two- and three-loop coefficients can be found in
refs.[\ref{Jones}--\ref{Tarasov}].

\subsection 3.2 Gauge fixing

The action $S[A]$ and the a priori measure $\rmD[A]$
are invariant under
arbitrary gauge transformations $A\to A^\Omega$
[cf.~eq.(2.40)].
We are interested in evaluating the functional integral
for fixed boundary values $\bvalue$ and $\bvalue'$,
and so, in this subsection,
restrict attention to the group
$\largegaugegroup$ of transformations $\Omega$
which leave the boundary values intact.
Our assumptions on the boundary values
imply that
a gauge function $\Omega$ belongs to $\largegaugegroup$
if and only if
$$
  \Omega(x)=\cases{
            \rme^{i2\pi m/N}  & at $x^0=0$,\cr
            \noalign{\vskip1ex}
            \rme^{i2\pi m'/N} & at $x^0=T$,\cr}
  \eqno\enum
$$
for some integers $m$ and $m'$.
$\largegaugegroup$ accordingly decomposes into $N^2$ disconnected
components.

The center $Z_N$ of $\largegaugegroup$
consists of all gauge functions $\Omega=\rme^{i2\pi m/N}$.
These transformations act trivially on
gauge fields and so it is really the factor group
$\gaugegroup=\largegaugegroup/Z_N$
with which we are concerned
when fixing the gauge.
$\gaugegroup$ may be identified with the
$m'=0$ component of $\largegaugegroup$.
It acts freely on the space of
gauge fields, i.e.~$A^{\Omega}=A$ for some $A$ implies
$\Omega=1$.

The gauge fixing procedure is the usual one:
we add a gauge fixing term
to the action and include an integration over
Faddeev-Popov ghost fields $c$ and $\bar{c}$ with the
appropriate action.
To write down the gauge fixing term,
the gauge field
$A$ integrated over is decomposed according to
$$
  A_{\mu}(x)=\bfield_{\mu}(x)+g_0\qfield_{\mu}(x).
  \eqno\enum
$$
The ``quantum" field $\qfield$ introduced here
is the new integration
variable, while $\bfield$ will be kept fixed.
In perturbation theory, only the
winding number zero sector contributes.
The boundary conditions
(2.20) thus become
$$
  \qfield_{k}(x)=0\quad\hbox{at $x^0=0$ and $x^0=T$}.
  \eqno\enum
$$
Note that the time component $\qfield_0$
remains unrestricted at this point.

In terms of the quantum field $\qfield$,
a gauge transformation $A\to A^{\Omega}$
amounts to the substitution
$$
  \qfield_{\mu}(x)\to\Omega(x)\qfield_{\mu}(x)\Omega(x)^{-1}+
  g_0^{-1}\left[\bfield_{\mu}^{\Omega}(x)-\bfield_{\mu}(x)\right].
  \eqno\enum
$$
In particular, if we set
$$
  \Omega(x)=1-g_0\omega(x)+\rmO(g_0^2),
  \eqno\enum
$$
the transformation becomes
$$
  \qfield_{\mu}(x)\to\qfield_{\mu}(x)+D_{\mu}\omega(x)+\rmO(g_0),
  \qquad D_{\mu}=\partial_{\mu}+\Ad \bfield_{\mu}.
  \eqno\enum
$$
This suggests that we take
$$
  S_{\rm gf}[\bfield,q]=
  -\lambda_0
  \int\rmd^Dx\,\tr\left\{
  D_{\mu}\qfield_{\mu}D_{\nu}\qfield_{\nu}\right\}
  \eqno\enum
$$
as the gauge fixing term, where
$\lambda_0$ denotes the bare gauge fixing parameter.
This gauge is referred to as the background gauge.
The associated Faddeev-Popov action reads
$$
  S_{\FP}[\bfield,\qfield,c,\bar{c}]=
  2\int\rmd^Dx\,\tr\left\{
  \bar{c}\,D_{\mu}(D_{\mu}+g_0\Ad \qfield_{\mu})c\right\},
  \eqno\enum
$$
and so we end up with
$$
  \rme^{-\effaction[\bfield]}=
  \int\rmD[\qfield]\int\rmD[c]\rmD[\bar{c}]\,
  \rme^{-S_{\rm total}[\bfield,\qfield,c,\bar{c}]},
  \eqno\enum
$$
where
$$
  S_{\rm total}[\bfield,\qfield,c,\bar{c}]=
  S[\bfield+g_0\qfield]+S_{\rm gf}[\bfield,\qfield]+
  S_{\rm FP}[\bfield,\qfield,c,\bar{c}]
  \eqno\enum
$$
is the total action.

We have already noted above that
the spatial components of $\qfield$ must vanish
at $x^0=0$ and $x^0=T$. The boundary conditions
on $\qfield_0$ and the Faddeev-Popov fields $c$ and $\bar{c}$
are determined by the gauge fixing procedure.
In particular, they depend on the choice of
the gauge fixing term
and on the gauge transformation
properties of the boundary values $\bvalue$ and $\bvalue'$.
The influence of the latter is subtle and will only become clear
after a while.

In the background gauge and for boundary values as specified
at the beginning of this section,
it turns out that
$$
  D_0\qfield_0(x)=c(x)=\bar{c}(x)=0
  \quad\hbox{at $x^0=0$ and $x^0=T$}
  \eqno\enum
$$
are the correct boundary conditions.
This is not easy to prove,
because the functional integral
is not really well-defined before we fix the gauge.
A rigorous derivation can (and will) be given
after we pass to the lattice formulation of the theory
(cf.~sect.~6).

At this point
it is still possible to explain, on a heuristic level,
how the boundary conditions (3.15) arise.
Let us first consider the Faddeev-Popov fields.
When going through the gauge fixing procedure,
one notes that the Faddeev-Popov operator
$$
  \Delta_{\FP}=-D_{\mu}\left(D_{\mu}+g_0\Ad\qfield_{\mu}\right)
  \eqno\enum
$$
acts in the linear
space of infinitesimal gauge transformations.
So if $t$ is an anti-commuting
parameter, the transformation
$\Omega(x)=\exp[tc(x)]$ must be an element of $\gaugegroup$,
and the same is true for $\bar{c}$.
{}From eq.(3.5) we now infer that the ghost fields
vanish at $x^0=0$ and $x^0=T$.

The boundary conditions on $\qfield_0$ are harder to justify.
An obvious case is the background field
Landau gauge $D_{\mu}\qfield_{\mu}=0$,
where the vanishing of $D_0\qfield_0$ at
the boundaries is an immediate consequence
of the boundary conditions on $\qfield_k$.
For general values of the gauge fixing parameter,
we may resort to the slightly obscure argument
that the gauge fixing function
$D_{\mu}\qfield_{\mu}$ should be a mapping from the space
of gauge fields to the space of infinitesimal gauge transformations.
Since the latter vanish at $x^0=0$ and $x^0=T$,
we again conclude that $\qfield_0$ satisfies the
boundary conditions (3.15).

\subsection 3.3 One-loop effective action

It is now straightforward to expand the
total action and the gauge fixed
functional integral (3.13) in powers of $g_0$.
To second order we have
$$
  S_{\rm total}[\bfield,\qfield,c,\bar{c}]=
  S[\bfield]-2\int\rmd^Dx\,\tr\left\{\frac{1}{2}
  \qfield_{\mu}\,\deltaonehat\qfield_{\mu}+
  \bar{c}\,\deltazerohat c\right\}
  +\rmO(g_0),
  \eqno\enum
$$
where the operators $\deltaonehat$ and $\deltazerohat$
are defined through
$$
  \eqalignno{
  \deltaonehat\qfield_{\mu}&=
  -D_{\nu}D_{\nu}\qfield_{\mu}
  +(1-\lambda_0)D_{\mu}D_{\nu}\qfield_{\nu}
  -2[G_{\mu\nu},\qfield_{\nu}],
  &\enum\cr
  \noalign{\vskip1.5ex}
  \deltazerohat c&=-D_{\nu}D_{\nu}c.
  &\enum\cr}
$$
$\deltaonehat$ and $\deltazerohat$ act on fields
in $D$ dimensions and are to be distinguished
from
$$
  \deltaone =\left.\deltaonehat \right|_{D=4}
  \quad\hbox{and}\quad
  \deltazero=\left.\deltazerohat\right|_{D=4}.
  \eqno\enum
$$
Both are hermitean operators
relative to the obvious scalar products
in the spaces of fields satisfying the boundary conditions
(3.7) and (3.15).
They are also elliptic (if $\lambda_0>0$)
and bounded from below.
In particular, there exists a complete set of eigenfunctions
with a discrete spectrum of real eigenvalues.

{}From the definition (3.19) it follows
that all eigenvalues
of $\deltazerohat$ are strictly positive.
Negative eigenvalues of $\deltaonehat$ are excluded too,
because $\bfield$ is a
minimal action configuration.
There could be a finite number of zero modes,
but this is an unlikely case which we shall not
discuss any further,
i.e.~our assumptions
on $\bfield$ are supplemented accordingly.

If we now insert eq.(3.17) in the functional integral
and expand the integrand,
we obtain the series (2.38) for the effective action.
The leading term is given by eq.(2.39) as expected.
At one-loop order
we only need to perform
the Gaussian integrals over the quantum
field and the Faddeev-Popov ghosts and so end up with
$$
  \effaction_1[\bfield]=\frac{1}{2}\ln\det\deltaonehat-
  \ln\det\deltazerohat.
  \eqno\enum
$$
The determinants
occuring here
are defined according to
the rules of dimensional regularization
(cf.~sect.~3.3 of ref.[\ref{LueDimReg}]).

In the following it is convenient to set $\lambda_0=1$.
We are free to make this choice, because
$\effaction_1[\bfield]$ is
independent of the gauge fixing parameter.
A rigorous proof of this
statement could actually be given at this point, but
since the argument is a bit lengthy (and the result is expected
anyway), we do not present it here.

A more explicit expression for the determinants,
exhibiting the dependence on
the space-time dimensionality,
can be given in terms of the heat kernels of
$\deltaonehat$ and $\deltazerohat$.
Heat kernels have simple factorization properties, and if
we take these into account, one gets
[\ref{LueDimReg}]
$$
  \eqalignno{
  \ln\det\deltaonehat&=
  -\int_0^{\infty}{\rmd t\over t}\,
  \left(\Tr\,\rme^{-t\deltaSone}\right)^{-2\varepsilon}
  \left(\Tr\,\rme^{-t\deltaone}
  -2\varepsilon\,\Tr\,\rme^{-t\deltazero}\right),
  &\enum\cr
  \noalign{\vskip2ex}
  \ln\det\deltazerohat&=
  -\int_0^{\infty}{\rmd t\over t}\,
  \left(\Tr\,\rme^{-t\deltaSone}\right)^{-2\varepsilon}
  \Tr\,\rme^{-t\deltazero}.
  &\enum\cr}
$$
The first factor in these integrals
arises from the added dimensions. It involves
the Laplacian $\deltaSone$ on the circle and is
explicitly given by
$$
  \Tr\,\rme^{-t\deltaSone}=
  \sum_{n=-\infty}^{\infty}\rme^{-t(2\pi n/L)^2}.
  \eqno\enum
$$
The other
factors, $\Tr\,\rme^{-t\deltaone}$ and $\Tr\,\rme^{-t\deltazero}$,
are defined in four dimensions and are thus independent of
$\varepsilon$. Finally,
the term proportional to $\varepsilon$
in eq.(3.22) must be included because the quantum field
$\qfield$ has $D$ vector components rather than $4$.

An important point to note is that
the integrals over the
proper time $t$ are convergent for $\Re\varepsilon>2$.
At the upper end of the integration range, convergence is
guaranteed by the exponential fall-off of the integrand
($\deltaone$ and $\deltazero$ are positive).
For $t\to0$, on the other hand, the asymptotic behaviour of the
heat kernels is computable. This goes under the name
of the Seeley-DeWitt expansion and will be discussed below.
The result is that the functions integrated over
are proportional to
$t^{-3+\varepsilon}$ at small $t$ and so are
integrable for $\Re\varepsilon>2$.

\subsection 3.4 Seeley-DeWitt expansion

To evaluate the one-loop effective action near four dimensions,
the integrals (3.22) and (3.23) must be analytically continued
from $\Re\varepsilon>2$ to a region
including the origin $\varepsilon=0$.
The key result
which enables us to do so is
the Seeley-DeWitt expansion.
There exists an extensive
literature on the subject
(an incomplete list of references is [\ref{DeWitt}--\ref{AvOs}])
and we shall, therefore,
be rather brief.

Let $\Delta$ stand for one of the operators $\deltaone$ or
$\deltazero$. The Seeley-DeWitt expansion states that
$$
  \Tr\,\rme^{-t\Delta}
  \mathrel{\mathop\sim_{t\to0}}
  \seeleycoeff{2}{\Delta}\,t^{-2}+
  \seeleycoeff{3/2}{\Delta}\,t^{-3/2}+
  \seeleycoeff{1}{\Delta}\,t^{-1}+\ldots,
  \eqno\enum
$$
where the coefficients
$\seeleycoeff{j/2}{\Delta}$
are algebraically computable.
There are two different kinds of contributions, the volume
and the boundary terms.
Volume terms only occur for even $j$ and are equal to
a local polynomial in the
background field $\bfield$ and its derivatives
integrated over the space-time manifold.
The boundary terms are constructed similarly, except that
the integral is taken over the hyper-planes at $x^0=0$ and
$x^0=T$ (with possibly a sign difference between the two).
The dimension of the polynomials must be equal to $4-j$
and $3-j$, respectively.

A crucial observation now is that
the spectrum of $\deltaone$ is invariant under
the substitution $\bfield\to\bfield^{\Omega}$,
for any gauge function $\Omega$.
We simply transform the eigenfunctions
covariantly, viz.
$$
  \qfield_{\mu}(x)\to\Omega(x)\qfield_{\mu}(x)\Omega(x)^{-1},
  \eqno\enum
$$
and then note that the eigenvalue equation and
the boundary conditions are preserved.
The same remark also applies to
$\deltazero$ and so it is clear that the coefficients
$\seeleycoeff{j/2}{\Delta}$ must be gauge invariant.

The gauge symmetry just described is not accidental.
The complete gauge fixed functional integral is actually
invariant if $\bfield$ is gauge transformed and the fields
integrated over are rotated covariantly. In particular,
as we have noted earlier, the effective action satisfies
$$
  \effaction[\bfield^{\Omega}]=
  \effaction[\bfield].
  \eqno\enum
$$
We emphasize that this property
does not depend on our choice of gauge fixing term.
The effective action is independent of the latter.
The advantage of the background gauge only is that
every individual contribution to the effective action
is already invariant, while in a general gauge
this would not be the case.

Besides the trivial polynomial (the constant) there is no
gauge invariant polynomial in the background field
and its derivatives with dimension less than 4.
The polynomials with dimension 4
are linear combinations of
$\tr\left\{G_{\mu\nu}G_{\rho\sigma}\right\}$.
Taking this into account it is clear that
all coefficients
$\seeleycoeff{j/2}{\Delta}$ with $j>0$
are independent of the background field.
Furthermore,
using one of the known
computational techniques (ref.[\ref{DuOlPe}] for example),
one finds that
$$
  \seeleycoeff{0}{\deltazero}=
  -\frac{1}{20}\seeleycoeff{0}{\deltaone}=
  {N\over96\pi^2}
  \int\rmd^4x\,\tr\left\{G_{\mu\nu}G_{\mu\nu}\right\}.
  \eqno\enum
$$
It has certainly not escaped the reader's notice that
our argumentation here parallels the discussion of the
renormalization of the Schr\"odinger functional in subsect.~2.5.
In particular, it is because of gauge invariance that
we are able to exclude the presence of
boundary terms in the coefficients $\seeleycoeff{0}{\Delta}$.
As we shall see below, this is intimately related to the fact
that the renormalization of the
effective action does not require any extra counterterms.

\subsection 3.5 Removal of the ultra-violet cutoff

To perform
the analytic continuation of the
integrals (3.22) and (3.23) towards $\varepsilon=0$,
it is useful to break up the integration range at $t=1$.
Since the integrands fall off exponentially, it is clear that
the integrals from 1 to infinity are
entire analytic functions of $\varepsilon$.
So we only need to care about the integration from 0 to 1.
In that range
the integrands may be split into singular and regular parts
according to
$$
  \Tr\,\rme^{-t\Delta}=\sum_{j=0}^4
  \seeleycoeff{j/2}{\Delta}\,t^{-j/2}
  +R(t|\Delta).
  \eqno\enum
$$
{}From the Seeley-DeWitt expansion we know that
$R(t|\Delta)$
is of order $t^{1/2}$ at small $t$.
Furthermore, using the Poisson summation formula, one may show
that
$$
  \Tr\,\rme^{-t\deltaSone}=
  {L\over(4\pi t)^{1/2}}\sum_{n=-\infty}^{\infty}
  \rme^{-n^2L^2/4t},
  \eqno\enum
$$
i.e.~this factor is proportional to $t^{-1/2}$.
The contributions of the
regular parts to the
integrals are hence
analytic in the region $\Re\varepsilon>-\frac{1}{2}$.
The remaining terms are easy to work out,
since the integrands are explicitly known.

To be able to write
the result of these computations concisely,
we introduce
the zeta function
$$
  \zetafunc{s}{\Delta}=\Tr\,\Delta^{-s}.
  \eqno\enum
$$
The trace converges for
$\Re s>2$, but after passing to the
heat kernel representation, one may show,
following the steps taken above,
that $\zetafunc{s}{\Delta}$ extends to a meromorphic function
in the whole complex plane.
In particular, there is no singularity at $s=0$
and we may define
$$
  \zetaprime{\Delta}=
  \left.{\rmd\over\rmd s}\zetafunc{s}{\Delta}\right|_{s=0}.
  \eqno\enum
$$
The determinants of $\deltaonehat$ and $\deltazerohat$
are now given by
$$
  \eqalignno{
  \ln\det\deltaonehat&=
  -\left[{1\over\varepsilon}+\ln4\pi
  -\euler+\frac{1}{10}\right]
  L^{-2\varepsilon}\seeleycoeff{0}{\deltaone}
  -\zetaprime{\deltaone},
  &\enum\cr
  \noalign{\vskip2ex}
  \ln\det\deltazerohat&=
  -\left[{1\over\varepsilon}+\ln4\pi-\euler\right]
  L^{-2\varepsilon}\seeleycoeff{0}{\deltazero}
  -\zetaprime{\deltazero},
  &\enum\cr}
$$
where $\euler=0.577...$ denotes Euler's constant
and all terms vanishing at $\varepsilon=0$ have been dropped.

{}From these results we immediately deduce that
$$
  \effaction_1[\bfield]
  \mathrel{\mathop=_{\varepsilon\to0}}
  -{11\over3\varepsilon}{N\over16\pi^2}
  \,\effaction_0[\bfield]
  +\rmO(1).
  \eqno\enum
$$
This singularity is exactly cancelled by the
coupling constant renormalization. Indeed, from eq.(3.2)
we get
$$
  \effaction[\bfield]=
  \mu^{-2\varepsilon}
  \left\{{1\over\gms^2}-z_1(\varepsilon)\right\}
  \effaction_0[\bfield]+
  \effaction_1[\bfield]
  +\rmO(\gms^2),
  \eqno\enum
$$
and after inserting eq.(3.4), we are left with
$$
  \eqalign{
  \left\{\effaction[\bfield]\right\}_{D=4}=
  &\left\{{1\over\gms^2}-
  {11\over3}{N\over16\pi^2}
  \left[\ln4\pi\mu^2-\euler+\frac{1}{11}\right]\right\}
  \effaction_0[\bfield]\cr
  \noalign{\vskip2ex}
  &-\frac{1}{2}\zetaprime{\deltaone}+\zetaprime{\deltazero}
  +\rmO(\gms^2).\cr}
  \eqno\enum
$$
We have thus
shown that to one-loop order
the Schr\"odinger functional renormalizes in the expected way.

The zeta functions appearing in eq.(3.37) are
complicated functionals of the background field, which
cannot normally be reduced to simpler expressions.
For spatially constant background fields
they can be
worked out numerically, essentially by listing all
eigenvalues of $\deltaone$ and $\deltazero$ up to a certain
size (cf.~sect.~7).


\section 4. Lattice Formulation

A non-perturbatively meaningful definition of the Schr\"odinger
functional can be given in the framework of lattice gauge theories.
The lattice regularization is not unique ---
a choice of lattice action must be made and some further
arbitrariness is involved when
the Schr\"odinger functional is introduced.
We expect that
these details do not matter in the continuum limit
and what follows should therefore be regarded
as one possible approach to the problem.

\subsection 4.1 Gauge fields

We choose to set up the lattice theory in euclidean space and thus
imagine that a regular hypercubic lattice
is superimposed on the space-time manifold.
$T$ is assumed to be an integer multiple of the lattice
spacing $a$ so that
the possible values of the time coordinate $x^0$
of a lattice point $x$
are $0,a,2a,\ldots,T$.

A lattice gauge field $U$
is an assignment of a link variable
$U(x,\mu)\in\SU$ to every pair
$(x,x+a\hat{\mu})$ of lattice points
($\hat{\mu}$ denotes the unit vector in
the $\mu$--direction).
In particular, the temporal link variables $U(x,0)$ are
defined for all lattice points with $0\leq x^0<T$.
Gauge functions $\Omega(x)$ live on the sites $x$ of the
lattice and take values in $\SU$. They
act on the link variables according to
$$
  U(x,\mu)\to U^{\Omega}(x,\mu)=\Omega(x)U(x,\mu)
  \Omega(x+a\hat{\mu})^{-1}.
  \eqno\enum
$$
As in the continuum we
require that
gauge fields and gauge transformation functions
are periodic in spatial directions with period $L$
(which must also be an integer multiple of $a$).

The lattice regularized Schr\"odinger functional
will be defined later on as an integral over all
lattice gauge fields with prescribed boundary values
$$
  \bvaluelat({\bf x},k)=\left.U(x,k)\right|_{x^0=0}
  \quad\hbox{and}\quad
  \bvaluelat'({\bf x},k)=\left.U(x,k)\right|_{x^0=T}.
  \eqno\enum
$$
To make contact with the continuum definition of the
Schr\"odinger functional, $\bvaluelat$ and $\bvaluelat'$
must be related to the continuum boundary
values $\bvalue$ and $\bvalue'$.

A natural relationship is suggested if we recall that
$U(x,\mu)$
is the parallel transporter for colour vectors
from $x+a\hat{\mu}$ to $x$ along the straight line connecting
these two points.
To achieve the best possible
matching of lattice and continuum boundary values,
we should thus identify
$\bvaluelat({\bf x},k)$
with the corresponding parallel transporter
determined by the field $\bvalue_k({\bf x})$.
In other words, we set
$$
  \bvaluelat({\bf x},k)=
  {\cal P}\exp
  \left\{a\int_0^1\rmd t\,
  \bvalue_k({\bf x}+a{\bf\hat{k}}-ta{\bf\hat{k}})\right\},
  \eqno\enum
$$
and $\bvaluelat'({\bf x},k)$ is similarly
given by the field $\bvalue'_k({\bf x})$
[in eq.(4.3) the symbol $\cal P$ implies
a path ordered exponential
such that the fields at the larger values of
the integration variable $t$ come first].
Note that
this construction is gauge covariant: if we
perform a
gauge transformation $\bvalue\to\bvalue^{\Lambda}$,
the associated boundary field
$\bvaluelat$ transforms as a lattice gauge field should.

\subsection 4.2 Lattice action and the Schr\"odinger functional

Following Wilson [\ref{Wilson}]
the action $S[U]$ of a lattice gauge field
$U$ is taken to be
$$
  S[U]=
  {1\over g_0^2}\sum_p w(p)\,
  \tr\left\{1-U(p)\right\},
  \eqno\enum
$$
where the sum runs over all
{\it oriented} plaquettes $p$
on the lattice and $U(p)$ denotes the parallel transporter
around $p$.
The weight $w(p)$ is equal to 1 in all cases
except for the spatial plaquettes
at $x^0=0$ and $x^0=T$
which are given the weight ${1\over2}$.
The significance of this weight factor
will be discussed later on.

The lattice regularized Schr\"odinger functional
is now defined through
$$
  \schrodinger[\bvalue',\bvalue]=
  \int\rmD[U]\,\rme^{-S[U]},
  \qquad
  \rmD[U]=\prod_{x,\mu}\rmd U(x,\mu),
  \eqno\enum
$$
where one integrates over all lattice gauge fields $U$ with
fixed boundary values as described above
[recall that $\rmd U$ denotes the normalized
invariant measure on $\SU$;
the product in eq.(4.5)
is over all lattice points $x$ and directions $\mu$
such that the link $(x,x+a\hat{\mu})$ is in the interior of the lattice].
We emphasize that the lattice Schr\"odinger functional
is regarded as a functional of the
continuous fields $\bvalue$ and $\bvalue'$
rather than
the boundary values of the link field.
The latter are
determined through eq.(4.3) and its primed relative.
In the continuum limit,
$\bvalue$ and $\bvalue'$ are kept fixed,
while the lattice spacing $a$ is sent to zero and
the bare coupling $g_0$ is scaled
according to the renormalization group.

Compared to the formal continuum expression (2.19) we seem to miss
a sum over topological classes here.
Such an average is
not needed on the lattice, because
the functional integral (4.5) is
already invariant
under {\it arbitrary} gauge transformations
of the boundary fields.
To see this, recall that
a gauge transformation
$\bvalue\to\bvalue^{\Lambda}$
induces a corresponding transformation
$\bvaluelat\to\bvaluelat^{\Lambda}$
of the lattice boundary field at $x^0=0$.
In the functional integral,
such a change of the boundary values can be compensated
by a substitution $U\to U^{\Omega}$
of the integration variables,
where $\Omega$ is an arbitrary gauge
transformation function with
$$
  \Omega(x)=\cases{\Lambda({\bf x}) & at $x^0=0$,\cr
  \noalign{\vskip1ex}
  1 & at $x^0=T$. \cr}
  \eqno\enum
$$
The crucial point to note is that
gauge functions with these
boundary values
exist on the lattice,
independently of whether $\Lambda$
is topologically trivial or not.

A quantum mechanical interpretation of
the lattice Schr\"odinger functional (4.5) can be
given through the well-known transfer matrix construction
[\ref{WilsonKogut}--\ref{LueLesHouches}].
The transfer matrix $\trans_0$ in the temporal gauge
$U(x,0)=1$ is a hermitean operator which acts
on Schr\"odinger wave functions $\psi[\bvaluelat]$
at time $x^0=0$.
It may be regarded as the lattice expression for
the step evolution operator $\rme^{-\ham a}$
\footnote{$\dag$}{\footnotefont
In some of the papers quoted above it is
the gauge projected
operator $\trans=\trans_0\projector$
which is referred to
as the transfer matrix.
There is no difference between $\trans$ and $\trans_0$
on the physical subspace. We here stick to $\trans_0$ to
match with the notation employed in sect.~2.}.
As in the continuum theory,
physical wave functions $\psi[\bvaluelat]$
must be gauge invariant. The associated
projector is again denoted by $\projector$ and
we may also introduce a state $|\bvaluelat\rangle$ in
a way analogous to $|\bvalue\rangle$
[cf.~eq.(2.9)].

The point we wish to make is that our definition of the
lattice Schr\"odinger functional is precisely
such that the identity
$$
  \schrodinger[\bvalue',\bvalue]=
  \langle\bvaluelat'|
  \left(\trans_0\right)^{T/a}\projector|\bvaluelat\rangle
  \eqno\enum
$$
holds.
This formula is entirely analogous to the corresponding
continuum expression (2.10). In particular, a spectral representation
of the type (2.11) may be obtained,
for all values of the lattice spacing.

\subsection 4.3 Background field

For given boundary values $\bvalue$ and $\bvalue'$
(with properties as specified earlier) and sufficiently
small lattice spacings, we expect that the
absolute minimum of the lattice action $S[U]$
is non-degenerate up to gauge transformations.
The minimizing configuration $\bfieldlat$
should moreover converge to the induced background field
$\bfield$ of the continuum theory,
$$
  \bfieldlat(x,\mu)=1+a\bfield_{\mu}(x)+\rmO(a^2),
  \eqno\enum
$$
provided that
both fields are transformed to a definite gauge
(the temporal gauge, for example).
In perturbation theory,
the lattice background field $\bfieldlat$
plays the same r\^ole as $\bfield$ did in the continuum.
It is, therefore,
often necessary to find an analytical or
at least an accurate numerical
representation of $\bfieldlat$.

To illustrate these remarks,
let us consider the self-dual
background field
which was introduced in sect.~2
[eqs.(2.32)--(2.36)].
Since $\bvalue$ and $\bvalue'$ are
spatially constant in this case,
the lattice boundary fields are simply given by
$$
  \bvaluelat({\bf x},k)=\exp\left\{a\bfunc(0)I_k\right\}
  \quad\hbox{and}\quad
  \bvaluelat'({\bf x},k)=\exp\left\{a\bfunc(T)I_k\right\}.
  \eqno\enum
$$
The associated minimal action configuration
$\bfieldlat$ must be a solution of the field equations
which one derives from the Wilson action $S[U]$.
To be able to write them concisely,
we introduce the plaquette field
$$
  \plaq(x,\mu,\nu)=
  \bfieldlat(x,\mu)\bfieldlat(x+a\hat{\mu},\nu)
  \bfieldlat(x+a\hat{\nu},\mu)^{-1}\bfieldlat(x,\nu)^{-1}
  \eqno\enum
$$
and its covariant divergence
$$
  \eqalignno{
  &\dplaq(x,\mu)=&\cr
  \noalign{\vskip1ex}
  &\sum_{\nu=0}^3\left\{
  \plaq(x,\mu,\nu)-
  \bfieldlat(x-a\hat{\nu},\nu)^{-1}\plaq(x-a\hat{\nu},\mu,\nu)
  \bfieldlat(x-a\hat{\nu},\nu)\right\}.&\enum\cr}
$$
One may now show that the lattice action is stationary
if and only if
the traceless anti-hermitean part of
$\dplaq(x,\mu)$ vanishes,
$$
  \dplaq(x,\mu)-\dplaq(x,\mu)^{\dagger}
  -{1\over N}\,\tr\left\{\dplaq(x,\mu)-
  \dplaq(x,\mu)^{\dagger}\right\}=0,
  \eqno\enum
$$
for all links $(x,x+a\hat{\mu})$ in the interior of the lattice.

The form
of the continuum background field suggests that we try
to solve the lattice field equations (4.12) with the ansatz
$$
  \bfieldlat(x,0)=1,\qquad
  \bfieldlat(x,k)=\exp\left\{a\bfunclat(x^0)I_k\right\},
  \eqno\enum
$$
where $\bfunclat$ is some real function
to be determined.
In the case of the $\SUtwo$ theory,
a solution is in fact obtained in this way,
if $\bfunclat$ satisfies
$$
  \partial^*\left[{1\over a^2}
  \sin\left(\frac{1}{2}a^2\partial\bfunclat\right)\right]
  =8\cos\left(\frac{1}{2}a\bfunclat\right)
  \left[{1\over a}
  \sin\left(\frac{1}{2}a\bfunclat\right)\right]^3,
  \qquad 0<x^0<T.
  \eqno\enum
$$
The operator $\partial$ in this equation denotes
the forward lattice derivative,
$$
  \partial f(x^0)=
  {1\over a}\left[f(x^0+a)-f(x^0)\right],
  \eqno\enum
$$
and $\partial^*$ the backward derivative.
The ansatz (4.13) also works out
for $\SUthree$,
but for $N\geq4$ we suspect that a more complicated
expression is needed, with
at least two unknown functions.

In the practically
relevant range of $T/a$, say $T/a=10,\ldots,100$,
the solution of eq.(4.14) with the required boundary values
(and the lowest action) can be determined
numerically to any desired precision.
Alternatively, if we assume that an expansion of the form
$$
  \bfunclat(x^0)=
  \bfunclat_0(x^0)+a\bfunclat_1(x^0)+a^2\bfunclat_2(x^0)+\ldots,
  \eqno\enum
$$
holds,
where $\bfunclat_k(x^0)$ are smooth functions independent of $a$,
the resulting tower of equations,
$$
  \eqalignno{
  \partial_0^2\bfunclat_0&=2\bfunclat_0^3,
  &\enum\cr
  \noalign{\vskip1ex}
  \partial_0^2\bfunclat_1&=6\bfunclat_0^2\bfunclat_1,
  &\enum\cr}
$$
etc., can be solved iteratively.
We expect, of course,
that the leading term $\bfunclat_0$ coincides
with the continuum solution $\bfunc$. This is
consistent with the lowest order equation (4.17),
and from eq.(4.18) we now infer that
$\bfunclat_1$ is a linear combination
of $\bfunc^2$ and $\bfunc^{-3}$.
If we take into account that $\bfunclat_1$ must vanish
at $x^0=0$ and $x^0=T$, a little calculation shows that
$\bfunclat_1$ is in fact equal to zero everywhere,
and so we conclude that
$$
  \bfunclat(x^0)=\bfunc(x^0)+\rmO(a^2).
  \eqno\enum
$$
We have also worked out the order $a^2$ correction
and compared our result with
the numerical solution of eq.(4.14).
Complete agreement was found, and
the approach of the lattice solution
to the continuum background field is, therefore,
well under control.

One point, however, that remains to be checked is
that the lattice background field so
constructed is indeed a configuration with
least action.
To convince oneself that there are no other
lower minima,
one may run a relaxation program,
on a range of lattices, starting from
various initial configurations.
In the case of most interest to us,
the constant
Abelian fields,
such a numerical ``proof\thinspace" of stability is fortunately not
needed,
because the minimum property of these fields
can be established by analytical means (cf.~sect.~5).

\subsection 4.4 Continuum limit in perturbation theory

The effective action on the lattice,
$\effaction[\bfield]$,
is defined through eq.(2.37)
and is considered
to be a functional of the continuum
background field.
We are certainly free to do so, because
the Schr\"odinger functional (4.5)
depends on the boundary
values $\bvalue$ and $\bvalue'$ and these are in
one-to-one correspondence with $\bfield$.
The notation is also in accord with our understanding
of the lattice as a device to regularize the theory
which is to be removed at fixed $\bfield$.

The perturbation expansion (2.38)
of the lattice effective action may
be derived following the steps taken in
sect.~3 for the case of dimensional regularization.
In particular, the leading term
$\effaction_0[\bfield]$ is equal to $g_0^2$ times
the action of the lattice background field $\bfieldlat$.
An interesting new aspect is that the gauge fixing
procedure can be carried out rigorously. This will
be discussed in sect.~6 and an explicit one-loop calculation
is presented in sect.~7.
Our aim here is to describe
how the continuum limit is reached in perturbation theory.

Symanzik's analysis
[\ref{SymanzikLELa},\ref{SymanzikLELb}]
of the cutoff dependence
of Feynman diagrams on the lattice suggests that
the $l$--loop contribution to the effective action may be expanded in
an asymptotic series of the form
$$
  \effaction_l[\bfield]
  \mathrel{\mathop\sim_{a\to0}}
  \sum_{m=0}^{\infty}\sum_{n=0}^{l}
  \effaction_{lmn}[\bfield]\,
  a^m(\ln a)^n.
  \eqno\enum
$$
Close to the continuum limit, all terms with $m>0$
can be neglected. We moreover expect that
the logarithmically divergent terms cancel after
the gauge coupling $g_0$ is renormalized,
i.e.~after we substitute
$$
  g_0^2=\glat^2+z_1(a\mu)\glat^4+\ldots,
  \qquad
  z_1(a\mu)={22\over3}{N\over16\pi^2}\ln(a\mu),
  \eqno\enum
$$
where $\glat$ is a renormalized coupling and $\mu$
some normalization mass.
This is entirely analogous to what happened in the
case of the dimensionally
regularized effective action.
In particular, the resulting renormalized expansion
in powers of $\glat^2$
must coincide with eq.(3.37), if the gauge couplings $\glat$
and $\gms$ are properly matched.

To one-loop order and for the background fields considered,
all these statements will be confirmed explicitly, and so
we are confident that the dependence of
the effective action on the lattice spacing
is indeed as described.

\subsection 4.5 $\rmO(a)$ improved action

Numerical simulations of the functional integral (4.5)
are limited to lattice spacings $a$ that are not too small
compared to the scales set by the background field.
The cutoff dependent contributions to the effective action
(the terms of order $a$ and higher) may hence have
a non-negligible influence on the outcome of such calculations.
In that instance
it may be desirable to choose an
improved lattice action so as to cancel the dominant
cutoff effects.

This idea has previously been worked out for scalar theories
[\ref{SymanzikImpA}--\ref{Keller}] and
pure gauge theories on lattices without boundaries
[\ref{WeiszImp}--\ref{LueWeImpB}].
The general principle is that the cutoff effects
of order $a(\ln a)^n$, $a^2(\ln a)^n$, etc.~can
be successively removed by adding a linear combination of
local counterterms
to the action, with increasing dimensions
and properly adjusted coefficients.
This may be viewed
as an extension of the renormalization
procedure to the level of
irrelevant operators.
We thus expect that the counterterms needed to
improve the Schr\"odinger functional
come in two forms, the
boundary and the volume terms (cf.~sect.~2).

At order $a$ there are two possible counterterms.
They are obtained by summing any local
lattice expression for the fields
$$
  a^4\,\tr\left\{F_{0k}F_{0k}\right\}
  \quad\hbox{and}\quad
  a^4\,\tr\left\{F_{kl}F_{kl}\right\}
  \eqno\enum
$$
over the $x^0=0$ and $x^0=T$ hyper-planes.
A simple $\rmO(a)$ improved
lattice action is thus given by eq.(4.4), where
$$
  w(p)=\cases{
  \frac{1}{2}c_s(g_0)
  & if $p$ is a spatial plaquette at $x^0=0$ or $x^0=T$, \cr
  \noalign{\vskip1ex}
  c_t(g_0)
  & if $p$ is a time-like plaquette attached to a  \cr
  & boundary plane,                                \cr}
  \eqno\enum
$$
and $w(p)=1$ in all other cases.
The coefficients
$c_s(g_0)$ and $c_t(g_0)$ multiplying the boundary
plaquettes allow us to monitor the
size of the counterterms. They
are to be adjusted so as to achieve the
desired improvement of the theory.

Ideally one would like to determine
$c_s(g_0)$ and $c_t(g_0)$ through numerical simulations
of the Schr\"odinger functional.
While this is not an impossible task, it is surely
demanding, since one needs
precise data on a range of lattices to be able to
isolate the cutoff effects properly.
In perturbation theory we have
$$
  \eqalignno{
  c_s(g_0)&=c_s^{(0)}+c_s^{(1)}g_0^2+\ldots,
  &\enum\cr
  \noalign{\vskip2ex}
  c_t(g_0)&=c_t^{(0)}+c_t^{(1)}g_0^2+\ldots,
  &\enum\cr}
$$
where the coefficients
$c_s^{(l)}$ and $c_t^{(l)}$ can be extracted from the
$l$--loop contribution to the effective action.
In particular, at tree level we only need to
work out the order $a$ term in the small $a$ expansion
of $\effaction_0[\bfield]$
for two independent choices of $\bfield$, say
a constant Abelian field and a self-dual field.
As a result one obtains
$$
  c_s^{(0)}=c_t^{(0)}=1\quad\hbox{for all $N$.}
  \eqno\enum
$$
On the basis of
our calculations to one-loop order (sect.~7),
we have moreover been able show that
$$
  c_s^{(1)}=-0.166(1)
  \quad\hbox{and}\quad
  c_t^{(1)}=-0.0543(5)
  \quad
  \hbox{for $N=2$.}
  \eqno\enum
$$
Note that in the $\SUtwo$ theory
the crossover from strong to weak coupling behaviour
occurs around $g_0^2=2$
and simulations nowadays are performed at couplings close to $1.5$.
The one-loop correction (4.27) is thus reasonably small.

We finally mention that the improved action
coincides with the Wilson action to lowest order of $g_0^2$.
The latter is hence $\rmO(a)$ improved at tree level,
and this implies that all coefficients
$\effaction_{lmn}[\bfield]$ in the expansion (4.20) with
$m=1$ and $n=l$ vanish.


\section 5. Abelian Background Fields

As explained in sect.~2, a large variety of
running couplings can be defined
by differentiating the effective action $\effaction[\bfield]$
with respect to some parameter
of the background field $\bfield$.
A careful choice of $\bfield$
is however necessary, if one attempts to study the
scaling properties of the coupling through numerical simulations
of the lattice Schr\"odinger functional.
In particular, since
one cannot afford to make the
lattice spacing arbitrarily small,
a background field is required for which
the lattice corrections to the effective action are tolerable.
It is our experience
that constant Abelian fields are optimal in this respect
and so we discuss their properties here.

In the following we allow
the space-time dimensionality $D$ to assume any integer value
greater or equal to 2. For the lattice action we take
Wilson's action (subsect.~4.2) multiplied with
$a^{D-4}$.

\subsection 5.1 Definition

The boundary values of the background fields
considered in this section
are spatially constant and diagonal. In other words, we have
$$
  \bvalue_{k}={i\over L}
  \pmatrix{\angle_{k1}  & 0            & \ldots & 0            \cr
           0            & \angle_{k2}  & \ldots & 0            \cr
           \vdots       & \vdots       & \ddots & \vdots       \cr
           0            & 0            & \ldots & \angle_{kN}  \cr},
  \eqno\enum
$$
where the angles
$\angle_{k\alpha}$ satisfy
$$
  \sum_{\alpha=1}^N\angle_{k\alpha}=0
  \quad\hbox{for all $k$.}
  \eqno\enum
$$
The other boundary field $\bvalue'$ is defined similarly,
with $\angle_{k\alpha}$
replaced by a second set of angles $\angleprime_{k\alpha}$.

Gauge fields of this type have previously occurred in the literature
and are referred to as ``torons"
(see e.g.~refs.[\ref{GJK},\ref{LueTorons}]).
Locally they are pure gauge configurations,
i.e.~the only gauge invariant information is contained
in the parallel transporters for closed curves winding
``around the world". In the field $\bvalue$ these are
products of the matrices $\exp\left\{L\bvalue_k\right\}$
and their inverses.
In particular, they only depend on the phase factors
$\rme^{i\angle_{k\alpha}}$ and
the angular character
of the parameters $\angle_{k\alpha}$ thus becomes apparent.

An obvious solution of the field equations with
the boundary values specified above is
$$
  \bfield_{0}=0,\qquad
  \bfield_k=\left[x^0\bvalue'_k
  +(T-x^0)\bvalue^{\vphantom{\prime}}_k\right]/T.
  \eqno\enum
$$
For the associated field tensor one obtains
$$
  \bfieldtensor_{0k}=\left[\bvalue'_k
  -\bvalue^{\vphantom{\prime}}_k\right]/T,
  \qquad\bfieldtensor_{kl}=0,
  \eqno\enum
$$
and the action is given by
$$
  S[\bfield]={L^{D-3}\over g_0^2T}
  \sum_{k=1}^{D-1}\sum_{\alpha=1}^N
  \left(\angleprime_{k\alpha}-\angle_{k\alpha}\right)^2.
  \eqno\enum
$$
This solution --- a constant colour electric field --- is
not the only solution
of the Yang-Mills field equations
with the required boundary values.
It is in fact quite obvious that other solutions must exist,
because the action (5.5) does not reflect the expected
periodicity in the angles
$\angle_{k\alpha}$.
The stability of the field (5.3) thus needs to be discussed
and we shall come back to this problem in subsect.~5.2.

The link field
$$
  \bfieldlat(x,\mu)=\exp\left\{a\bfield_{\mu}(x)\right\}
  \eqno\enum
$$
(with $\bfield$ as above) is a candidate for
the induced background field on the lattice.
$\bfieldlat$ has the required boundary values
and one may easily verify that it is a solution of
the lattice field equations (4.12).
Note that the action
$$
  S[\bfieldlat]={TL^{D-1}\over g_0^2}
  \sum_{k=1}^{D-1}\sum_{\alpha=1}^N
  \left\{{2\over a^2}\sin\left[{a^2\over2TL}
  \left(\angleprime_{k\alpha}-\angle_{k\alpha}\right)
  \right]\right\}^2
  \eqno\enum
$$
coincides with the continuum action (5.5) up to terms of
order $a^4$ (rather than $a^2$).
Constant Abelian fields thus show their quality of leading
to minimal cutoff effects already at the classical level.

\subsection 5.2 Stability theorem

Before the lattice field (5.6)
can be accepted as the induced
background field for the chosen boundary values,
we must show that it is a configuration with
least action and that any other field
with the same action is gauge equivalent to $\bfieldlat$.
Such a proof can in fact be given if
$\angle_{k\alpha}$ and $\angleprime_{k\alpha}$,
are restricted to a certain bounded domain.

To specify this region it is useful to arrange
$\angle_{k\alpha}$ (and similarly $\angleprime_{k\alpha}$) into
$D-1$ colour vectors $\anglevector_k$ in the obvious way.
We say that a vector
$\anglevector=(\angle_1,\ldots,\angle_N)$
is in the {\it fundamental domain}\/ if
$$
  \angle_1<\angle_2<\ldots<\angle_N, \qquad
  \left|\angle_{\alpha}-\angle_{\beta}\right|<2\pi,\qquad
  \sum_{\alpha=1}^N\angle_{\alpha}=0.
  \eqno\enum
$$
This is a bounded convex set, which has a certain
group theoretical significance. Namely, if $u\in\SU$
has pairwise different eigenvalues $\lambda_{\alpha}$,
there exists a unique ordering of the eigenvalues
and a unique vector
$\anglevector$ in the fundamental domain, such that
$\lambda_{\alpha}=\rme^{i\angle_{\alpha}}$.

Our result on the stability of the background fields considered
in this section is now summarized by

\proclaim
Theorem 1.
Suppose the angle vectors $\anglevector_k$ and
$\anglevectorprime_k$ are in the fundamental domain for
all $k=1,\ldots,D-1$. Let $\bfieldlat$ be the associated
background field (5.6) on a lattice with
$$
  TL/a^2>(N-1)\pi^2\max\{1,N/16\}.
  \eqno\enum
$$
The action of any other lattice gauge field $U$ with the
same boundary values then satisfies
$S[U]\geq S[\bfieldlat]$, where the equality holds
if and only if $U$ is gauge equivalent
to $\bfieldlat$.

\noindent
The proof of the theorem is quite involved and is deferred
to appendix B. The condition (5.9) is of a technical
nature and is perhaps not really needed for the validity
of the theorem. It is, in any case, insignificant
for $N\leq3$ and the lattices of interest.

There is, of course, a continuum version of the theorem,
provided one is willing to restrict
attention to the category of
differentiable fields. The minimum property
of the solution (5.3) is in fact a simple corollary
of theorem 1,
since any differentiable gauge field can be approximated
arbitrarily well by lattice fields.

\subsection 5.3 Definition of a running coupling
                for the $\SUtwo$ theory

We are now ready to introduce
the running coupling
$\gbar^2(L)$ which will be
used in our finite size scaling study
of the $\SUtwo$ theory in four dimensions
and which will, therefore, be in the focus of interest
for the rest of this paper.

Following the general
scheme described in subsect.~2.6,
we choose the solution (5.3) for the background field $\bfield$
and set $T=L$.
$\bfield$ depends on
altogether 6 parameters,
say $\angle_{k1}$ and $\angleprime_{k1}$.
A symmetrical one-parameter
submanifold is
$$
  \angle_{k1}=-\bfieldparm,\qquad\angleprime_{k1}=-\pi+\bfieldparm,
  \eqno\enum
$$
and the coupling is now given by eq.(2.43).

We should of course make sure that the premises of theorem 1
are satisfied so that the stability of the chosen background
field is guaranteed. This is the case if
$$
  0<\bfieldparm<\pi\quad\hbox{and}\quad L/a\geq4.
  \eqno\enum
$$
In the numerical work we take $\bfieldparm=\pi/4$,
which is half-way between
the zero action point $\bfieldparm=\pi/2$
and the boundary of the stability interval.
All these details are quite arbitrary,
but we emphasize (once more) that
they are in no way
of fundamental importance for
the scaling analysis we have in mind.


\section 6. Fixing the Gauge in the Lattice Theory

The problem of gauge fixing has already been addressed
in sect.~3 on a somewhat formal level.
In particular, a truly convincing derivation of the
boundary conditions on the Faddeev-Popov ghosts and
the time component of the gauge field could not be given and
so we feel that it is worthwhile to
go through the gauge fixing procedure once more,
in the mathematically rigorous framework of the lattice theory.
We closely follow the general
scheme discussed in subsect.~3.5 of
ref.[\ref{LueLesHouches}], where further details
and a proof of the basic lemma can be found.

In this section the background field is assumed to be
an Abelian field as described in sect.~5,
with angle vectors $\anglevector_k$ and $\anglevectorprime_k$
in the fundamental domain.

\subsection 6.1 Group of gauge transformations

The lattice Schr\"odinger functional (4.5) is invariant
under all gauge transformations $U\to U^{\Omega}$
which leave the boundary fields $\bvaluelat$ and $\bvaluelat'$
intact.
According to lemma 1 of appendix B,
this condition is satisfied if and only if
the boundary values
of the gauge function $\Omega(x)$ at $x^0=0$ and $x^0=T$
are constant and diagonal.
The group of all these transformation functions
is denoted by $\largegaugegroup$.

The constant diagonal gauge functions $\Omega(x)$ form
a subgroup of $\largegaugegroup$ isomorphic to
the Cartan subgroup $\cartangroup$ of $\SU$.
Such transformations act trivially on the background field
and there are actually no further transformations with this
property (cf.~appendix B).
The gauge directions in the space of infinitesimal
deformations of the background field are thus generated by
$$
  \gaugegroup=\largegaugegroup/\cartangroup,
  \eqno\enum
$$
and so it is this set of transformations which needs to be fixed.
The group $\cartangroup$ then survives as a global symmetry
of the theory.

In the following it is convenient to identify
$\gaugegroup$ with the group of all transformations
$\Omega\in\largegaugegroup$ that are equal to $1$ at $x^0=T$.
{}From a purely mathematical point of view, we then have
the canonical situation, with a compact gauge group
acting freely on the space of fields integrated over.
The discussion in
ref.[\ref{LueLesHouches}] thus carries over literally.

\subsection 6.2 Spaces of infinitesimal fields

The Lie algebra $\gaugealgebra$ of $\gaugegroup$
consists of all fields $\omega(x)$ such that
the infinitesimal transformation
$$
  \Omega(x)=1-g_0\omega(x)+\rmO(g_0^2)
  \eqno\enum
$$
belongs to $\gaugegroup$.
In other words,
$\omega$ must be a spatially
periodic lattice field which takes
values in the Lie algebra
of $\SU$ and which satisfies
the boundary conditions
$$
  \left.\omega(x)\right|_{x^0=0}=\kappa,
  \qquad
  \left.\omega(x)\right|_{x^0=T}=0,
  \eqno\enum
$$
where $\kappa$ is constant and diagonal.

A linear space $\fluctuationspace$ of lattice
vector fields $\qfield_{\mu}(x)$ may be introduced similarly.
We here require that
$$
  U(x,\mu)=\left\{1+g_0a\qfield_{\mu}(x)+\rmO(g_0^2)\right\}
  \bfieldlat(x,\mu)
  \eqno\enum
$$
is a valid infinitesimal deformation of the background field.
In particular, to guarantee that $U$ has the same boundary values
as $\bfieldlat$, the spatial components
$\qfield_k(x)$ must
be equal to zero at $x^0=0$ and $x^0=T$.

We shall later find it useful to have a scalar product
on these spaces at our disposal.
The obvious choice for the product of two vector fields is
$$
  (q,r)=-2a^4\sum_{x,\mu}
  \tr\left\{q_{\mu}(x)r_{\mu}(x)\right\},
  \eqno\enum
$$
and the product on $\gaugealgebra$ is defined similarly.

Another notational item are the covariant lattice derivatives
$\cdev{\mu}$ and $\cdevstar{\mu}$.
These operators act on functions
$f(x)$ with values in the Lie algebra of $\SU$
and are given by
$$
  \eqalignno{
  \cdev{\mu}f(x)&={1\over a}\left[
  \bfieldlat(x,\mu)f(x+a\hat{\mu})\bfieldlat(x,\mu)^{-1}
  -f(x)\right],
  &\enum\cr
  \noalign{\vskip1.5ex}
  \cdevstar{\mu}f(x)&={1\over a}\left[
  f(x)-
  \bfieldlat(x-a\hat{\mu},\mu)^{-1}
  f(x-a\hat{\mu})\bfieldlat(x-a\hat{\mu},\mu)\right].
  &\enum\cr}
$$
They
only make sense if $f$ is defined
at the lattice points referred to
and should thus be used with care.

\subsection 6.3 Gauge fixing function

The gauge fixing term to be added to the Wilson action
is the square of a suitable
gauge fixing function $\gaugefunction$.
As explained in
ref.[\ref{LueLesHouches}],
$\gaugefunction$ must be a regular
mapping from the space of fields integrated over
to the Lie algebra $\gaugealgebra$.
For perturbation theory
it is actually sufficient to specify
$\gaugefunction$ on an arbitrarily small but finite
neighborhood of the background field $\bfieldlat$.
The fields $U$ in such a neighborhood may be
parametrized by
$$
  U(x,\mu)=\exp\left\{g_0a\qfield_{\mu}(x)\right\}\,
  \bfieldlat(x,\mu),
  \eqno\enum
$$
where $\qfield\in\fluctuationspace$ and say
$\|\qfield\|<\varepsilon$.
The simplest possibility then is to choose
the gauge fixing function to be a
linear mapping from
$\fluctuationspace$ to $\gaugealgebra$.

To be able to write it in a compact form
we introduce the operator
$$
  d:\;\gaugealgebra\mapsto\fluctuationspace,
  \qquad
  \left(d\omega\right)_{\mu}(x)=\cdev{\mu}\omega(x).
  \eqno\enum
$$
The covariant derivative occurring here is well-defined,
for all relevant lattice points $x$ and directions $\mu$,
and it is also easy to verify that $d\omega$ has the proper
boundary values. The significance of $d$
becomes apparent if we note that the gauge directions in
the space of infinitesimal deformations of the background field
are precisely the modes of the form $d\omega$.
As already mentioned, the group $\gaugegroup$
acts freely on the space of gauge fields, and so
it is not surprising that the kernel of $d$
turns out to be trivial,
i.e.~$d\omega=0$ implies $\omega=0$.

The basic property the gauge fixing function should have is
that it does {\it not}\/ vanish on the gauge modes $d\omega$.
It is quite obvious now that
this requirement (and all the other conditions listed in
ref.[\ref{LueLesHouches}])
will be fulfilled if we choose
$$
  F(U)=d^{\ast}\qfield,
  \eqno\enum
$$
where $d^{\ast}$ is minus the adjoint of $d$.
That is, $d^{\ast}$ maps any vector field
$\qfield\in\fluctuationspace$ onto an element of $\gaugealgebra$
such that
$$
  (d^{\ast}\qfield,\omega)=-(\qfield,d\omega)
  \quad\hbox{for all}\quad\omega\in\gaugealgebra.
  \eqno\enum
$$
This property implies
$$
  d^{\ast}\qfield(x)=
  \cdevstar{\mu}\qfield_{\mu}(x)
  \quad\hbox{if}\quad 0<x^0<T,
  \eqno\enum
$$
and for the boundary values at
$x^0=0$ and $x^0=T$ one obtains
$$
  \left[d^*\qfield(x)\right]_{\alpha\beta}=
  \cases{
  (a^2/L^3)
  \sum_{\bf y}\left[\qfield_0(0,{\bf y})\right]_{\alpha\beta}
  & if $\alpha=\beta$ and $x^0=0$, \cr
  \noalign{\vskip1ex}
  0
  & otherwise. \cr}
  \eqno\enum
$$
Note that these are indeed as required for
a function contained in $\gaugealgebra$.

\subsection 6.4 Gauge fixed functional integral

Now that we have specified the
gauge fixing function $\gaugefunction$,
the gauge fixed form of the Schr\"odinger functional (4.5)
is obtained almost mechanically,
following the steps described
in ref.[\ref{LueLesHouches}].
There is no new element involved here and so we merely
introduce the necessary notations and state the
final result.

Our choice of gauge fixing function corresponds to
the gauge fixing term
$$
  S_{\rm gf}[\bfield,\qfield]=
  {\lambda_0\over2}(d^*\qfield,d^*\qfield).
  \eqno\enum
$$
In view of eq.(6.12), this may be regarded as a lattice
version of the background gauge.
The associated Faddeev-Popov ghosts $c$ and $\bar{c}$
are in all respects like infinitesimal gauge transformations
except that they are fermi fields.
That is, if we choose a complete
set of functions in $\gaugealgebra$,
the ghost fields are equal to
the general linear combination of these basis elements,
with anti-commuting coefficients that generate
a Grassmann algebra.

The action of the Faddeev-Popov fields is
$$
  S_{\FP}[\bfield,\qfield,c,\bar{c}]=
  -(\bar{c},d^*\delta_c\qfield),
  \eqno\enum
$$
where $\delta_{c}\qfield$
denotes the first order variation of $\qfield$
under the gauge transformation
generated by $c$.
To second order in $g_0$ we have
$$
  \eqalignno{
  \delta_{c}\qfield_{\mu}&=
  \cdev{\mu}c
  +g_0\Ad\qfield_{\mu}\,c
  &\cr
  \noalign{\vskip2.5ex}
  &\quad+\left[\frac{1}{2}g_0a\Ad \qfield_{\mu}+
  \frac{1}{12}\left(g_0a\Ad \qfield_{\mu}\right)^2+\ldots\right]
  \cdev{\mu}c
  &\enum\cr}
$$
(no sum over $\mu$ is implied here).
Note that $\delta_c\qfield$ is a vector field with the correct
boundary values. The action of $d^*$ in eq.(6.15) is therefore
well-defined.

The gauge fixed form of the Schr\"odinger functional
is now given by
$$
  \eqalignno{
  \rme^{-\effaction[\bfield]}&=
  \int\rmD[U]
  \int\rmD[c]\rmD[\bar{c}]
  \,\rme^{-S_{\rm total}[\bfield,\qfield,c,\bar{c}]},
  &\enum\cr
  \noalign{\vskip2ex}
  S_{\rm total}[\bfield,\qfield,c,\bar{c}]&=
  S[U]+
  S_{\rm gf}[\bfield,\qfield]+
  S_{\FP}[\bfield,\qfield,c,\bar{c}],
  &\enum\cr}
$$
where it is understood that $U$
and $\qfield$ are related by eq.(6.8).
The first integral in eq.(6.17)
is restricted to a small neighborhood of the
background field $\bfieldlat$, i.e.~we are neglecting
terms that are exponentially small in the coupling.
The integration over the ghost fields is the usual one,
resulting in the determinant of the Faddeev-Popov operator.

As in the continuum theory, the gauge fixed functional
integral is the starting point from which
the perturbation expansion of the effective action
is obtained.
Noting
$$
  \eqalignno{
  \rmD[U]&=\rmD[q]\left\{1+\rmO(g_0^2)\right\},
  &\enum\cr
  \noalign{\vskip1.5ex}
  S_{\rm total}[\bfield,\qfield,c,\bar{c}]
  &=S[\bfieldlat]+
  \frac{1}{2}\left(\qfield,\deltaone\qfield\right)+
  \left(\bar{c},\deltazero c\right)+\rmO(g_0),
  &\enum\cr}
$$
we get, for the first two terms in eq.(2.38),
$$
  \eqalignno{
  \effaction_0[\bfield]&=
  g_0^2S[\bfieldlat],
  &\enum\cr
  \noalign{\vskip1.5ex}
  \effaction_1[\bfield]&=
  \frac{1}{2}\ln\det\deltaone-\ln\det\deltazero.
  &\enum\cr}
$$
$\deltazero$ and $\deltaone$
are symmetric operators
acting on functions
in $\gaugealgebra$ and $\fluctuationspace$ respectively.
We have already used the symbols $\deltazero$ and $\deltaone$
for the corresponding operators in the continuum theory,
but this will not lead any confusion in the following.

{}From the above we infer that
$$
  \eqalignno{
  \deltazero&=-d^*d,
  &\enum\cr
  \noalign{\vskip1ex}
  \deltaone&=\deltaoneprime-\lambda_0dd^*,
  &\enum\cr}
$$
where $\deltaoneprime$ is obtained by expanding
the Wilson action $S[U]$ to second order in $\qfield$.
With the help of the
$\star$ product notation introduced in appendix A,
the result of that computation
may be written in the compact form
$$
  \eqalignno{
  &\deltaoneprime\qfield_{\mu}(x)=
  \sum_{\nu\neq\mu}\Bigl\{
  \cosh\left(a^2\bfieldtensor_{\mu\nu}\right)\star
  \left[-\cdevstar{\nu}\cdev{\nu}\qfield_{\mu}(x)
        +\cdevstar{\nu}\cdev{\mu}\qfield_{\nu}(x)\right]
  &\enum\cr
  \noalign{\nobreak\vskip1.5ex}
  &-a^{-2}\sinh\left(a^2\bfieldtensor_{\mu\nu}\right)\star
  \left[2\qfield_{\nu}(x)
        +a\left(\cdevstar{\nu}+\cdev{\nu}\right)\qfield_{\mu}(x)
        +a^2\cdevstar{\nu}\cdev{\mu}\qfield_{\nu}(x)\right]
  \Bigr\}.
  &\cr}
$$
It should be emphasized that
this formula is
only valid for constant Abelian background fields,
the case of immediate interest to us.
We have also
assumed that the link $(x,x+a\hat{\mu})$ is
contained in the interior of the lattice, viz.
$$
  \hbox{$0\leq x^0<T$
  if $\mu=0$ and $0<x^0<T$ if $\mu>0$.}
  \eqno\enum
$$
All degrees of freedom of the fields in $\fluctuationspace$
reside on these links and the operator $\deltaoneprime$
is thus completely specified by eq.(6.25).

\subsection 6.5 Boundary conditions

The lattice quantum field $\qfield$ and the
ghost fields $c$ and $\bar{c}$ do not satisfy any boundary
conditions beyond those implicit in the definition of
$\fluctuationspace$ and $\gaugealgebra$.
In particular, the time component
$\qfield_0(x)$ is defined at all points
$x$ with $0\leq x^0<T$ and is unconstrained.

However, after passing to the continuum limit,
the situation regarding
boundary conditions may be quite different.
In a free scalar theory, for example, the propagator
on a lattice with free boundary conditions converges to
a Green function which satisfies Neumann boundary conditions.
We should thus be prepared to find that the
eigenfunctions of the lattice operators $\deltazero$
and $\deltaone$ satisfy additional boundary conditions
in the limit $a\to0$.

To be absolutely clear,
we are not suggesting that the lattice theory should be
amended in some way.
Our aim rather is to determine
which boundary conditions must be chosen
when one attempts to formulate
the theory directly in the continuum, using dimensional
regularization for example.
By comparing with the lattice theory
this question can now be decided,
because
all cutoff prescriptions should
yield identical results at tree level.

In the case of the lattice operator $\deltaone$
it is not immediately obvious
how to pass to the continuum limit, because
the definition of the gauge term
$dd^*$ is slightly non-uniform at the boundary
[cf.~eqs.(6.12) and (6.13)].
This difficulty can be removed by choosing
a better notation. To this end
we extend the time component $\qfield_0(x)$ of the lattice field
to all points with $x^0=-a$ and $x^0=T$. Its values there
are chosen such that
$$
  \cdevstar{0}\qfield_{0}(x)=d^*\qfield(x)
  \quad\hbox{at $x^0=0$ and $x^0=T$}.
  \eqno\enum
$$
No additional degrees of freedom are thus introduced and
the extension should be regarded purely as a matter of
convention.

The operator $dd^*$ is now given by the simple expression
$$
  dd^*\qfield_{\mu}(x)=\cdev{\mu}\cdevstar{\nu}\qfield_{\nu}(x),
  \eqno\enum
$$
for all $x$ and $\mu$ in the range (6.26).
Furthermore, we may interpret
eq.(6.27) as a boundary condition
on $\qfield_0$. At $x^0=T$, for example,
the requirement simply is that
$\cdevstar{0}\qfield_0(x)=0$.
At the other end of the lattice
$\qfield_0$ is again required to satisfy
Neumann boundary conditions with
the exception of
the spatially constant diagonal modes, which
must vanish at $x^0=-a$ and
thus satisfy Dirichlet boundary conditions.
This latter complication depends on
our choice of background field and is absent
for the irreducible background fields considered in sect.~3.

In the present formulation one may straightforwardly
pass to the continuum limit.
In particular,
taking $a\to0$ in eqs.(6.25) and (6.28)
leads to the continuum expression (3.18)
for $\deltaone$
and the boundary conditions
stated above carry over literally.
To remove all doubts about this procedure, we have
verified in a number of cases, by exact numerical computation,
that the eigenfunctions
of the lattice operator converge to smooth functions
in the continuum limit and that these indeed
satisfy the expected boundary conditions.

We thus conclude that the proper
boundary conditions
on the quantum field $\qfield$
in the continuum theory are a mixture of Dirichlet and
Neumann boundary conditions
as described above.
A similar analysis applies to
the Faddeev-Popov fields. The result here is that besides
the boundary conditions already present on the lattice,
the spatially constant diagonal modes
satisfy Neumann boundary conditions at $x^0=0$.
This is again special to our choice of background field in this section,
i.e.~for irreducible boundary values $\bvalue$ and $\bvalue'$
one would end up with Dirichlet boundary conditions for all
modes.


\section 7. First Order Correction to the Running Coupling
[$\SUtwo$ Theory]

We now set $N=2$
and consider the one-parameter
family of background fields defined
in subsect.~5.3.
Our aim is to compute
the associated running coupling
$\gbar^2(L)$ to one-loop order of
perturbation theory.
The calculation will be performed
in the framework of the lattice
theory, but
as a check we have also worked out
the relevant determinants
in the continuum theory,
using dimensional regularization. Some
of the details involved in that calculation will
be sketched at the end of this section.

\subsection 7.1 Preliminaries

In the following we choose lattice units
and thus set $a=1$.
{}From the perturbation expansion of the effective action
we deduce that
$$
  \gbar^2(L)=g_0^2+m_1(L)g_0^4+m_2(L)g_0^6+\ldots
  \eqno\enum
$$
The one-loop coefficient
is a sum of two contributions,
$$
  m_1=h_0-\frac{1}{2}h_1,
  \eqno\enum
$$
one from the ghost and the other from the vector fields.
Explicitly, they are given by
$$
  h_s=
  {\partial\over\partial\bfieldparm}
  \ln\det\Delta_s
  \Bigm/
  {\partial\over\partial\bfieldparm}
  \effaction_0[\bfield],
  \qquad s=0,1.
  \eqno\enum
$$
By differentiating the classical action (5.7) one obtains
$$
  {\partial\over\partial\bfieldparm}
  \effaction_0[\bfield]=
  -24L^2\sin\left[{1\over L^2}(\pi-2\bfieldparm)\right],
  \eqno\enum
$$
and we are thus left with the problem to compute the
determinants of $\deltazero$ and $\deltaone$
for the background fields of interest.
We do not expect that this can be done analytically,
but as will be shown in the following subsections,
it is possible to evaluate the determinants numerically,
for a useful
range of $L$ and to essentially any desired precision.

\subsection 7.2 Symmetries

$\deltazero$ is a symmetric operator which acts
in a real vector space. In particular, there exists
a complete set of eigenfunctions and the determinant
of $\deltazero$ is the product of all eigenvalues.

In the following we consider
$\deltazero$ to be a linear
operator in the complex space
$\gaugealgebra\oplus i\gaugealgebra$.
The eigenfunctions and eigenvalues are not affected
by this extension and so it is clear that also
the determinant is the same as before.
The reason for going to the complexified space is
that now we can more easily pass to a basis
which is adapted to the symmetries of the problem.

To exploit the invariance of $\deltazero$
under constant diagonal gauge transformations,
we choose the standard basis
$T^a$ of $\SUtwo$ generators [eq.(A.4)] and define
$T^{\pm}=T^1\pm iT^2$.
We then consider the decomposition
$$
  \gaugealgebra\oplus i\gaugealgebra=
  \ga{0}\oplus\ga{-}\oplus\ga{+}
  \eqno\enum
$$
where
$\ga{0}$ and $\ga{\pm}$ are the subspaces
of all fields that are proportional to $T^3$ and
$T^{\pm}$ respectively.
Constant diagonal gauge transformations reduce
to a multiplication by a phase factor on $\ga{\sigma}$
and so we expect that these spaces are invariant under
the action of $\deltazero$.

To show this explicitly, we introduce the corresponding
subspaces $\fs{0}$ and $\fs{\pm}$ of complex vector fields
and note that the operator $d$
maps any function $\omega\in\ga{\sigma}$
onto an element of $\fs{\sigma}$.
More precisely, we have
$$
  \cdev{\mu}\omega(x)=
  \cases{
  \omega(x+\hat{\mu})-\omega(x)
  & if $\mu=0$,  \cr
  \noalign{\vskip1.5ex}
  \rme^{i\sigma\beta(x^0)}\omega(x+\hat{\mu})-\omega(x)
  & if $\mu>0$,  \cr}
  \eqno\enum
$$
where the phase $\beta$ is given by
$$
  \beta(t)=-(2/L^2)
  \left[\bfieldparm L+(\pi-2\bfieldparm)t\right].
  \eqno\enum
$$
It is quite obvious now that
$d^*$ maps $\fs{\sigma}$ into $\ga{\sigma}$
and so we conclude that $\deltazero$ leaves
$\ga{\sigma}$ invariant.

The operator $\deltaone$ is similarly reduced by
the subspaces $\fs{\sigma}$. This is an immediate consequence
of the identities
$$
  \eqalignno{
  \cosh\bfieldtensor_{0k}\star T^a
  &=\cos(\gamma/2)T^a,
  \qquad
  \gamma=(2/L^2)(\pi-2\bfieldparm),
  &\enum\cr
  \noalign{\vskip1.5ex}
  \sinh\bfieldtensor_{0k}\star T^3
  &=0,
  &\enum\cr
  \noalign{\vskip1.5ex}
  \sinh\bfieldtensor_{0k}\star T^{\pm}
  &=\mp i\sin(\gamma/2)T^{\pm},
  &\enum\cr}
$$
and the discussion above of the covariant derivatives.
In particular, we have
$$
  \det\deltaone=
  \det\left.\deltaone\right|_{\fs{0}}
  \det\left.\deltaone\right|_{\fs{-}}
  \det\left.\deltaone\right|_{\fs{+}},
  \eqno\enum
$$
and the analogous factorization holds for the determinant
of $\deltazero$. Note that the last two factors
in eq.(7.11) are equal.

A further factorization of the determinants is obtained
if we take into account that
$\deltazero$ and $\deltaone$
are invariant under spatial translations.
The eigenfunctions of this group of symmetries
are the plane
waves $\rme^{i{\bf px}}$ where
$$
  {\bf p}=2\pi{\bf n}/L,
  \qquad
  n_k\in\gz,
  \qquad
  -L/2<n_k\leq L/2.
  \eqno\enum
$$
We are thus led to introduce
the subspaces $\gap{\sigma}$ and $\fsp{\sigma}$ of
all functions
which are proportional to
$\rme^{i{\bf px}}$ and have no other dependence on $\bf x$.
Since the momentum ${\bf p}$ is conserved under the action of
$\deltaone$, it follows that
$$
  \det\left.\deltaone\right|_{\fs{\sigma}}
  =\prod_{\bf p}
  \det\left.\deltaone\right|_{\fsp{\sigma}},
  \eqno\enum
$$
and the same factorization holds in the case of the
operator $\deltazero$.

The simplification which has thus been achieved is substantial.
In each symmetry sector the spatial
dependence of the wave functions
and also the $\SUtwo$ degrees of freedom are completely fixed.
The problem is, therefore,
effectively reduced to computing the determinants of a
set of finite difference operators in one dimension.

\subsection 7.3 Computation of $h_0$

{}From the discussion above
it is evident that $\deltazero$
is independent of the background field on the subspace $\ga{0}$.
The contribution from this sector hence drops out and
we are left with
$$
  {\partial\over\partial\bfieldparm}
  \ln\det\deltazero=
  2\sum_{\bf p}
  {\partial\over\partial\bfieldparm}
  \ln\det\left.\deltazero\right|_{\gap{+}}.
  \eqno\enum
$$
So let us consider the action of $\deltazero$ in the subspace
$\gap{+}$ in some more detail.
The functions in this symmetry sector are of the form
$$
  \omega(x)=
  \psi(x^0)\,\rme^{i{\bf px}}\,T^{+},
  \eqno\enum
$$
where $\psi(t)$, $0\leq t\leq T$,
is a complex function with
$$
  \psi(0)=\psi(T)=0
  \eqno\enum
$$
[cf.~eq.(6.3); the constant $\kappa$ vanishes because
$T^{+}$ is off-diagonal].

On this set of functions
$\deltazero$ reduces to an ordinary
second order difference operator,
$$
  \deltazero\psi(t)=
  \amat(t)\psi(t+1)+\bmat(t)\psi(t)+\cmat(t)\psi(t-1),
  \eqno\enum
$$
with coefficients given by
$$
  \eqalignno{
  \amat(t)&=\cmat(t)=-1,
  &\enum\cr
  \noalign{\vskip1ex}
  \bmat(t)&=
  8-2\sum_{k=1}^3\cos\left[p_k+\beta(t)\right].
  &\enum\cr}
$$
The determinant of such operators can be computed
by solving a simple recursion relation.
This is a known trick which
is discussed in some generality
in appendix C.
In the present case we need to solve
$$
  \deltazero\psi(t)=0,\qquad 0<t<T,
  \eqno\enum
$$
starting from
the initial values
$$
  \psi(0)=0,\qquad\psi(1)=1.
  \eqno\enum
$$
The determinant is then given by
$$
  \det\left.\deltazero\right|_{\gap{+}}=\psi(T).
  \eqno\enum
$$
Note that the calculational effort to solve eq.(7.20)
is proportional to $T$, while one usually needs of order $T^3$
arithmetic operations to
evaluate the determinant of
a $T\times T$ matrix.

To compute the derivative of the determinant with respect
to $\bfieldparm$, the best way to proceed is to
derive a recursion for $\partial\psi(t)/\partial\bfieldparm$
by differentiating eq.(7.20).
Both recursion relations are trivial to program,
and so one is able to compute the coefficient $h_0$
for say $L\leq32$ with a negligible amount of computer time.

\subsection 7.4 Computation of $h_1$

For notational convenience
we always assume that
the field $\qfield_0(x)$ is extended one step beyond the lattice
to all points with $x^0=-1$ and $x^0=T$,
as discussed in subsect.~6.5.
The boundary conditions on $\qfield_{\mu}$
then are a mixture of Dirichlet and
Neumann boundary conditions and the action
of $\deltaone$ is given by eqs.(6.24),(6.25) and (6.28).

We first consider the contribution of the sector $\fsp{0}$.
The fields in this subspace may be written as
$$
  \eqalignno{
  \qfield_{0}(x)&=
  \psi_0(x^0)(-i)\rme^{i{\bf px}}\,T^3,
  &\enum\cr
  \noalign{\vskip1.5ex}
  \qfield_{k}(x)&=
  \psi_k(x^0)\rme^{{i\over2}p_k}\rme^{i{\bf px}}\,T^3,
  &\enum\cr}
$$
where $\psi_{\mu}(t)$ is some complex vector field.
The extra phase factors in these equations have no particular
significance except that they lead to simpler expressions later on.

The action of the operator $\deltaone$ on wave functions
of the above type is of the general form
$$
  \deltaone\psi_{\mu}(t)=
   \amat_{\mu\nu}(t)\psi_{\nu}(t+1)
  +\bmat_{\mu\nu}(t)\psi_{\nu}(t)
  +\cmat_{\mu\nu}(t)\psi_{\nu}(t-1).
  \eqno\enum
$$
It is not difficult to work out the coefficient matrices
$\amat$, $\bmat$ and $\cmat$ explicitly [cf.~appendix D].

For all momenta $\bf p\neq0$ the boundary conditions
on $\psi_{\mu}(t)$ are
$$
  \partial^*\psi_0(t)=\psi_k(t)=0
  \quad\hbox{at $t=0$ and $t=T$}.
  \eqno\enum
$$
$\partial^*$ here denotes
the backward difference, defined in the usual way.
If $\bf p=0$
it is $\psi_0(-1)$ instead of
$\partial^*\psi_0(0)$ which is required to vanish,
while all other boundary conditions are as above.

The determinant of $\deltaone$ in this latter sector
can actually be computed immediately.
The matrices $\amat$, $\bmat$ and
$\cmat$ assume a particularly simple form in this case,
leading to a decoupling of the $\psi_0$ and $\psi_k$ components.
Furthermore, only the spatial modes make a field dependent
contribution and one quickly finds that
$$
  {\partial\over\partial\bfieldparm}
  \ln\det\left.\deltaone\right|_{\fs{0}({\bf 0})}=
  3(T-1)
  {\partial\over\partial\bfieldparm}
  \ln\cgh.
  \eqno\enum
$$
This is nice because now we are left with the $\bf p\neq0$
subspaces, where the boundary conditions are simply given by
eq.(7.26).

To compute the associated determinants, we proceed
essentially as in the case of the Faddeev-Popov operator.
We first construct all solutions of
$$
  \eqalignno{
  \deltaone\psi_0(t)&=0,
  \qquad0\leq t<T,
  &\enum\cr
  \noalign{\vskip1ex}
  \deltaone\psi_k(t)&=0,
  \qquad0<t<T,
  &\enum\cr}
$$
with $\partial^*\psi_0(0)=\psi_k(0)=0$.
For specified
initial values,
$\psi_0(0)$ and $\partial\psi_k(0)$,
there exists exactly one solution. It is
obtained recursively, first by solving eq.(7.28) at $t=0$,
then eq.(7.29) at $t=1$, and so on.

It is useful to group the initial data
in a vector $v_{\mu}$ with components
$$
  v_0=\psi_0(0),\qquad v_k=\partial\psi_k(0).
  \eqno\enum
$$
After completing the recursion we may compute
the final values
$$
  w_0=\partial^*\psi_0(T),\qquad w_k=\psi_k(T).
  \eqno\enum
$$
It is evident that these are related linearly to the
initial values, i.e.~there exists a
matrix $\solmat$
such that
$$
  w_{\mu}=\solmat_{\mu\nu}v_{\nu}.
  \eqno\enum
$$
The determinant of $\deltaone$ is then given by
$$
  \det\left.\deltaone\right|_{\fsp{0}}=
  \lambda_0^T\left(\cgh\right)^{3T-3}\det\solmat,
  \eqno\enum
$$
as one may show by adapting the arguments of appendix C
to the case at hand.

Let us now turn to the subspaces $\fsp{+}$.
The wave functions here are of the form
$$
  \eqalignno{
  \qfield_{0}(x)&=
  \psi_0(x^0)(-i)\rme^{i{\bf px}}\,T^+,
  &\enum\cr
  \noalign{\vskip1.5ex}
  \qfield_{k}(x)&=
  \psi_k(x^0)\rme^{{i\over2}\left[p_k+\beta(x^0)\right]}
  \rme^{i{\bf px}}\,T^+,
  &\enum\cr}
$$
and the action of $\deltaone$ is again given by
eq.(7.25), with coefficients $\amat$, $\bmat$ and
$\cmat$ listed in appendix D.
Furthermore,
the boundary conditions are as before [eq.(7.26)],
and so it is clear that the determinant of $\deltaone$
in this sector can be computed following the lines
discussed above.
In particular, if $M$ is the matrix
connecting initial and final values, we have
$$
  \det\left.\deltaone\right|_{\fsp{+}}=
  \lambda_0^T\det\solmat,
  \eqno\enum
$$
for all momenta $\bf p$.


\topinsert
\newdimen\digitwidth
\setbox0=\hbox{\rm 0}
\digitwidth=\wd0
\catcode`@=\active
\def@{\kern\digitwidth}
\tablecaption{
One-loop coefficient
$m_1(L)$ and remainder $\varepsilon(L)$ at $\bfieldparm=\pi/4$
}
$$\vbox{\settabs\+  xx$10$
                 &  xxx$1.00000000$
                 &  xxx$1.00000$
                 &  xxxxxx$10$
                 &  xxx$1.00000000$
                 &  xxx$1.00000$
                 &  xx\cr
\vskip0.5ex
\thicktablerule
\vskip1ex
                \+  \hfill $L$              \hskip0.0ex
                 &  \hfill $m_1(L)$         \hskip2.0ex
                 &  \hfill $\varepsilon(L)$ \hskip1.0ex
                 &  \hfill $L$              \hskip0.0ex
                 &  \hfill $m_1(L)$         \hskip2.0ex
                 &  \hfill $\varepsilon(L)$ \hskip1.0ex
                 &  \quad\cr
\vskip 0.5ex
\thintablerule
\vskip 1ex
  \+ \hfill$ 6$ & \hfill$0.35422030$ & \hfill$0.00352$ &
     \hfill$20$ & \hfill$0.47541031$ & \hfill$0.00024$ & \cr
  \+ \hfill$ 8$ & \hfill$0.38370429$ & \hfill$0.00177$ &
     \hfill$22$ & \hfill$0.48471069$ & \hfill$0.00020$ & \cr
  \+ \hfill$10$ & \hfill$0.40644167$ & \hfill$0.00107$ &
     \hfill$24$ & \hfill$0.49316916$ & \hfill$0.00017$ & \cr
  \+ \hfill$12$ & \hfill$0.42483107$ & \hfill$0.00072$ &
     \hfill$26$ & \hfill$0.50092488$ & \hfill$0.00014$ & \cr
  \+ \hfill$14$ & \hfill$0.44023578$ & \hfill$0.00052$ &
     \hfill$28$ & \hfill$0.50808526$ & \hfill$0.00012$ & \cr
  \+ \hfill$16$ & \hfill$0.45347745$ & \hfill$0.00039$ &
     \hfill$30$ & \hfill$0.51473490$ & \hfill$0.00010$ & \cr
  \+ \hfill$18$ & \hfill$0.46508332$ & \hfill$0.00031$ &
     \hfill$32$ & \hfill$0.52094162$ & \hfill$0.00009$ & \cr
\vskip 1ex
\thicktablerule}$$
\endinsert

\subsection 7.5 Results

We now set $\bfieldparm=\pi/4$
and compute the one-loop correction
to the running coupling
using the techniques described above.
Some of our results are listed in table 1.
The calculations
were done so as to guarantee a final precision
of at least 20 decimal places. In particular,
all digits quoted in table 1 are significant.

In lattice units
the continuum limit is approached by
taking $L$ to infinity.
As discussed in subsects.~4.4 and 4.5 we
expect that an asymptotic expansion of the form
$$
  m_1(L)
  \mathrel{\mathop\sim_{L\to\infty}}
  \sum_{n=0}^{\infty}
  \left(r_n+s_n\ln L\right)/L^n
  \eqno\enum
$$
holds, with
$$
  s_0={11\over12\pi^2}
  \quad\hbox{and}\quad
  s_1=0.
  \eqno\enum
$$
The first few coefficients $r_n$ and $s_n$
may be extracted from the data listed in table 1
by any suitable extrapolation procedure.
In particular, using the blocking method of
ref.[\ref{LueWeImpB}], we were able to confirm,
to four significant decimal places,
that $s_0$ is given by eq.(7.38).
This again shows that all ultra-violet divergences
in the Schr\"odinger functional cancel after
renormalizing the coupling in the usual way.

If we now assume that $s_0$ is precisely given by eq.(7.38),
the extrapolation of our data yields
$$
  \eqalignno{
  r_0&=
  0.202349(3),
  &\enum\cr
  \noalign{\vskip1ex}
  r_1&=
  -0.1084(11).
  &\enum\cr}
$$
As shown by table 1, the remainder
$$
  \varepsilon(L)=
  m_1(L)-s_0\ln L-r_0-r_1/L,
  \eqno\enum
$$
with coefficients as given above, is rapidly decreasing
over the range of $L$ covered.
The higher order cutoff effects are thus comfortably small for
our choice of background field.
It must be emphasized, however, that other
background fields,
such as the constant self-dual field
considered in sect.~2, do not fare as well in this respect.
Partly this is because the strength
of the Abelian field is constant while in the other cases
the lattice must be fine enough
to resolve the often appreciable time dependence of the field.

After eliminating $g_0$ in favour of
the renormalized coupling $\glat$ introduced in
subsect.~4.4, the continuum limit can be taken
and one ends up with the expansion
$$
  \gbar^2(L)=\glat^2+
  \left[{11\over12\pi^2}\ln(\mu L)+r_0\right]\glat^4
  +\ldots
  \eqno\enum
$$
At this point we may pass to any other renormalization scheme
defined in the continuum theory.
In particular,
if we choose
the running coupling $\gbarms^2(L)$ in the $\ms$
scheme of dimensional regularization as our new expansion
parameter and use
the known relation between $\glat$ and $\gms$
[\ref{Hasenfratz},\ref{DashenGross},\ref{Weisz}],
we obtain
$$
  \gbar^2(L)=\gbarms^2(L)+c_1\gbarms^4(L)+\ldots,
  \eqno\enum
$$
where the coefficient $c_1$ is given by
$$
  c_1={11\over24\pi^2}\left[\ln4\pi-\euler-1.61638(8)\right].
  \eqno\enum
$$
For the corresponding $\Lambda$ parameter ratio one finds
$$
  \Lambda/\Lambda_{\ms}=1.18378(5),
  \eqno\enum
$$
which shows that
our coupling is very nearly equal to
$\gbarms^2(L)$ in the perturbative domain.

The term proportional to
$r_1$ in the expansion (7.37) is an order $a$ cutoff effect.
As discussed in subsect.~4.5 it can be removed by
adding a combination
of two boundary counterterms to the Wilson action,
with coefficients $c_s$ and $c_t$.
Our result above, eq.(7.40), allows us to compute $c_t$
to one-loop order [cf.~eq.(4.27)].

To also determine $c_s$,
the effective action of
a further
background field, with non-vanishing spatial field strength,
must be computed.
The constant self-dual fields are a convenient choice for
this calculation, because the determinants of $\deltazero$
and $\deltaone$ again factorize after going to momentum space.
With some additional work, we have thus been able
to obtain the result quoted in subsect.~4.5.

\subsection 7.6 Computation of\/ $\gbar^2(L)$ using dimensional
regularization

It is possible to derive the expansion (7.43)
directly in the continuum theory,
starting from the dimensionally regularized Schr\"odinger
functional. This provides an important check on our
calculations on the lattice and gives us additional confidence
that the Schr\"odinger functional is indeed
a universal amplitude.

In sect.~3 we have already obtained the one-loop effective
action for a general irreducible background field, using dimensional
regularization.
The final formula, eq.(3.37),
is also valid for Abelian background fields, if
$\deltazero$ and $\deltaone$ are considered to be operators
in the proper function spaces
[cf.~subsect.~6.5].
The non-trivial remaining problem then is
to compute
$\zetaprime{\deltazero}$ and $\zetaprime{\deltaone}$
or rather their derivatives with respect to
$\bfieldparm$.

In the following we adopt the notation of sect.~3.
In particular, the symbol $\Delta$ denotes any one of the
operators $\deltazero$ or $\deltaone$.
{}From the definition of the zeta function and the
Seeley-DeWitt expansion we infer that
$$
  \eqalignno{
  &\zetaprime{\Delta}=\euler\seeleycoeff{0}{\Delta}+
  &\cr
  \noalign{\vskip2ex}
  &\int_{0}^{\infty}{\rmd t\over t}\biggl\{
  \Tr\,\rme^{-t\Delta}-\sum_{j=1}^{4}
  \seeleycoeff{j/2}{\Delta}\,t^{-j/2}
  -\seeleycoeff{0}{\Delta}\theta(1-t)\biggr\}.
  &\enum\cr}
$$
This integral is absolutely convergent and
the lower end of the integration range
may, therefore,
be replaced by some positive cutoff $\delta$ which is
subsequently sent to zero.
It is then straightforward to show that
$$
  {\partial\over\partial\bfieldparm}
  \zetaprime{\Delta}=
  \lim_{\delta\to0}\left\{
  (\euler+\ln\delta)
  {\partial\over\partial\bfieldparm}
  \seeleycoeff{0}{\Delta}-
  \sum_{n=0}^{\infty}
  \rme^{-\delta E_n}
  {\partial\over\partial\bfieldparm}\ln E_n
  \right\},
  \eqno\enum
$$
where $E_n$, $n=0,1,2,\ldots$, are the eigenvalues of $\Delta$.

Our computational strategy now is as follows.
We first calculate all eigenvalues $E_n$ and their derivatives
$\partial E_n/\partial\bfieldparm$ up to a certain level.
For values of $\delta$ that are not too small,
the bracket in eq.(7.47) is then accurately given by
restricting the spectral sum to these eigenvalues.
Finally, the result is extrapolated to the limit $\delta=0$,
taking into account that the bracket has an asymptotic
expansion in powers of $\delta^{1/2}$.

The success of the method of course depends on
our ability to compute sufficiently many eigenvalues of $\Delta$.
This goal is easy to reach.
We first fix the symmetry
properties of the wave functions as before,
by choosing an appropriate basis of group generators and
by going to momentum space.
The range of momenta $\bf p$ is
here unbounded, but since the eigenvalues are growing
essentially like ${\bf p}^2$, only a finite subset of
momenta needs to be considered.

In every symmetry sector the problem is then reduced to
finding the eigenvalues of a certain
second order ordinary differential operator acting on functions
with Dirichlet or Neumann boundary conditions.
The reduced
operators are actually so simple that the eigenfunctions
can be given explicitly in terms of
hypergeometric functions with a parameter to be adjusted.
Alternatively, one may use exact numerical methods
(Runge-Kutta integration and root finding algorithms) or
one may set up a variational calculation with
a suitable basis of plane wave states.

Proceeding either way we have been able, with a small amount
of computer time, to calculate the bracket in eq.(7.47)
for values of $\delta$ in the range between say
$0.01\times L^2$ and $0.001\times L^2$.
{}From there an extrapolation to $\delta=0$ is safely possible
and our result for
the coefficient $c_1$ was found to be in agreement
with eq.(7.44)
(the estimated precision after the extrapolation
was about 4 significant digits).


\section 8. Concluding Remarks

The renormalizability of the Schr\"odinger functional
opens the door to a new generation of scaling studies of
pure gauge theories. We have in this paper prepared the
ground for such an investigation.
Numerical simulations of the $\SUtwo$ theory
are now underway and first results are expected soon.

It would be quite trivial to extend our
calculations in sect.~7
of the running coupling
to the physically more interesting case of the $\SUthree$
theory.
Before that one however needs
to fix the parameters of the background field.
In particular, some experience must be accumulated
to determine which fields lead to an acceptably large
signal in numerical simulations,
while keeping the cutoff effects low.

We also believe that an extension of our computations to
the two-loop level is feasible, although this is clearly
going to be hard work.


\appendix A

Our notational conventions are as follows.
Lorentz
indices $\mu,\nu,\ldots$ normally
run from 0 to 3. In the context of dimensional regularization
they run up to $D-1$, the dimension of space.
Since the metric
is euclidean, it does not matter in which position these indices appear.
The totally anti-symmetric symbol $\epsilon_{\mu\nu\rho\sigma}$
is normalized such that $\epsilon_{0123}=1$.
Latin indices $k,l,...$ run from 1 to 3
(or $D-1$) and are used to label spatial
vector components.
Colour vectors in the fundamental representation of $\SU$
carry indices $\alpha,\beta,\ldots$ ranging from 1 to $N$,
while for vectors in the adjoint representation, latin
indices $a,b,\ldots$
running from 1 to $N^2-1$ are employed.
Repeated indices are automatically summed over
unless stated otherwise.

The Lie algebra $\su$ of $\SU$
can be identified with the space
of complex $N\times N$ matrices
$X_{\alpha\beta}$ which satisfy
$$
  X^{\dagger}=-X,\qquad \tr\{X\}=0,
  \eqno\enum
$$
where $X^{\dagger}$ denotes the adjoint matrix
of $X$ and $\tr\{X\}=X_{\alpha\alpha}$
is the trace of $X$.
This is a {\it real} vector space of dimension
$N^2-1$. A natural inner product is given by
$$
  (X,Y)=-2\,\tr\{XY\},
  \eqno\enum
$$
and we may thus choose a basis
$T^a$, $a=1,2,\ldots,N^2-1$, such that
$$
  (T^a,T^b)=\delta^{ab}.
  \eqno\enum
$$
For $N=2$, the standard basis is
$$
  T^a={\pauli{a}\over2i},\quad a=1,2,3,
  \eqno\enum
$$
where $\pauli{a}$ denote the Pauli matrices.
With these conventions,
the structure constants $f^{abc}$, defined through
$$
  \left[T^a,T^b\right]=f^{abc}T^c,
  \eqno\enum
$$
are real and totally anti-symmetric under permutations of the
indices.

An $\SU$ gauge potential is a vector field
$A_{\mu}(x)$ with values in the Lie algebra $\su$.
With respect to a basis $T^a$ of group generators, we have
$$
  A_{\mu}(x)=A_{\mu}^{a}(x)T^a,
  \eqno\enum
$$
where the component fields $A_{\mu}^{a}(x)$ are real.
The components of
other fields such as the gauge field tensor
or the Faddeev-Popov ghost field
are defined in the same way.

The representation space of the adjoint representation of $\su$ is
the Lie algebra itself, i.e.~the
elements $X$ of $\su$ are represented by linear transformations
$$
  \Ad X:\quad\su\mapsto\su.
  \eqno\enum
$$
Explicitly, $\Ad X$ is defined through
$$
  \Ad X(Y)=[X,Y]\quad\hbox{for all $Y\in\su$.}
  \eqno\enum
$$
With respect to a basis $T^a$,
the associated matrix $(\Ad X)^{ab}$ representing the transformation is
given by
$$
  \Ad X(T^b)=T^a(\Ad X)^{ab},
  \eqno\enum
$$
which is equivalent to
$$
  (\Ad X)^{ab}=-f^{abc}X^c,\qquad
  \big(\Ad X(Y)\big)^a=f^{abc}X^bY^c,
  \eqno\enum
$$
in terms of the structure constants.

We finally define the $\star$ product
of two
$N\times N$ matrices $M$ and $X$ through
$$
  M\star X={1\over2}\left(MX+XM^{\dagger}\right)-
  {1\over2N}\,\tr\left(MX+XM^{\dagger}\right).
  \eqno\enum
$$
$M\star X$ is contained in $\su$,
for all $X\in\su$ and arbitrary $M$.


\appendix B

The proof of theorem 1 is elementary from the mathematical
point of view, but it is lengthy and is therefore broken up
into several steps. Our first goal will be to show that
the theorem holds in all dimensions if it is true in two
dimensions. After that we shall work out the consequences
of the lattice field equations in two dimensions.
We shall find that there is a discrete
set of solutions and we are then left to prove
that among these the configuration with least action
is the background field $\bfieldlat$.

Without further notice, we assume in the following
that the angle vectors $\anglevector_k$ and $\anglevectorprime_k$
are in the fundamental domain (5.8) and that the bound (5.9)
holds.
The discussion will be solely concerned with the lattice theory
and the boundary values of the gauge fields
considered are always as
specified in subsect.~5.1. Furthermore, the letter $\bfieldlat$
is strictly reserved for the background field defined through
eqs.(5.3) and (5.6).

\subsection B.1 Remarks on gauge transformations

The boundary fields $\bvaluelat$ and $\bvaluelat'$
have an important stability property under gauge transformations
which is expressed by

\proclaim
Lemma 1.
If $\Lambda$ is a spatial gauge transformation which leaves
$\bvaluelat$ or $\bvaluelat'$ invariant, it must be
equal to a constant diagonal matrix.

\proof
Let us assume that it is $\bvaluelat$ which is invariant
under the action of $\Lambda$, viz.
$$
  \Lambda({\bf x})\bvaluelat({\bf x},k)
  \Lambda({\bf x}+a{\bf\hat{k}})^{-1}=
  \bvaluelat({\bf x},k).
  \eqno\enum
$$
The parallel transporter along any lattice curve at $x^0=0$
is an ordered products of the link variables and so must be
invariant too. In particular, we may consider a closed
curve starting at some point $x$ and
winding once around the world in the $k$--direction.
Since $\Lambda$ is periodic, we then deduce that
$$
  \Lambda({\bf x})\exp\left\{L\bvalue_k\right\}\Lambda({\bf x})^{-1}
  =\exp\left\{L\bvalue_k\right\}.
  \eqno\enum
$$
The parallel transporter
$\exp\left\{L\bvalue_k\right\}$ is a diagonal matrix with eigenvalues
$\lambda_{\alpha}=\rme^{i\angle_{k\alpha}}$.
These are all distinct (the angle vector $\anglevector_k$ is in
the fundamental domain) and eq.(B.2) thus implies that
$\Lambda({\bf x})$ must be diagonal.
That it must also be constant,
is now an immediate consequence of eq.(B.1).
\endproof

A simple corollary of this lemma is that the background field
$\bfieldlat$ is invariant under a space-time gauge transformation
$\Omega$ if and only if $\Omega$ is constant and diagonal.

For any given lattice gauge field $U$ we define
$$
  \temptrans(x)=U(y,0)U(y+a\hat{0},0)U(y+2a\hat{0},0)\ldots
  U(x-a\hat{0},0),\qquad y=(0,{\bf x}),
  \eqno\enum
$$
which is just the parallel transporter along the time axis
from $x$ to the
``base point" $y$ at the boundary.
$\temptrans$ is periodic in the space directions
and may be regarded as a gauge transformation function.

\proclaim
Lemma 2. If $U$ is known to be gauge equivalent to $\bfieldlat$, we have
$\bfieldlat=U^{\temptrans}$.

\proof
Let us define $\widetilde{U}=U^{\temptrans}$. From the definition
(B.3) of $\temptrans$ we infer that
$$
  \eqalignno{
  \widetilde{U}(x,0)&=\bfieldlat(x,0)=1\quad\hbox{for all $x$,}
  &\enum\cr
  \noalign{\vskip1ex}
  \widetilde{U}(x,k)&=\bfieldlat(x,k)=\bvaluelat({\bf x},k)
  \quad\hbox{for all $k$ and all $x$ with $x^0=0$.}
  &\enum\cr}
$$
Furthermore, since $U$ and $\bfieldlat$
are gauge equivalent, there exists a gauge transformation $\Omega$
such that $\widetilde{U}=\bfieldlat^{\Omega}$.

Eq.(B.4) now implies that $\Omega$ is time independent,
and if we combine eq.(B.5) with lemma 1, one concludes
that $\Omega$ must be constant and diagonal. It follows
that $\bfieldlat^{\Omega}=\bfieldlat$ and so we have proved the lemma.
\endproof

\subsection B.2 Reduction to two dimensions

A significant simplification is now achieved by

\proclaim
Lemma 3. Let us assume that theorem 1 holds in two dimensions.
Then it is true for all $D\geq2$.

\proof
The action of any field $U$ in $D$ dimensions may be
written in the form
$$
  S[U]=S_E[U]+S_M[U],
  \eqno\enum
$$
where $S_E$ and $S_M$ are the contributions of the time-like
and spatial plaquettes, respectively.
The magnetic part $S_M$ is always greater or equal to zero
and so we conclude that
$$
  S[U]\geq S_E[U].
  \eqno\enum
$$
The electric part $S_E$ may be split up according to
$$
  S_E[U]=a^{D-2}\sum_{\cal P}
  S[\left.U\right|_{\cal P}].
  \eqno\enum
$$
The sum here extends over all two-dimensional time-like planes
$\cal P$ on the lattice and
$\left.U\right|_{\cal P}$
is the two-dimensional gauge field that one obtains by restricting
$U$ to $\cal P$.

We now make use of the premise of the lemma to show that
$$
  S_E[U]\geq a^{D-2}\sum_{\cal P}
  S[\left.\bfieldlat\right|_{\cal P}]
  =S_E[\bfieldlat]=S[\bfieldlat].
  \eqno\enum
$$
When combined with eq.(B.7), this
proves that $S[U]\geq S[\bfieldlat]$ for all fields $U$.

We still need to show that $U$ is gauge equivalent to $\bfieldlat$
if $S[U]=S[\bfieldlat]$. So let us assume that $U$ is a
minimal action configuration.
{}From the above we then infer that
the two-dimensional fields
$\left.U\right|_{\cal P}$ and $\left.\bfieldlat\right|_{\cal P}$
are gauge equivalent, for all planes $\cal P$.
This does not immediately imply that
$U$ is gauge equivalent to $\bfieldlat$, because
the gauge transformations which map
$\left.U\right|_{\cal P}$
onto $\left.\bfieldlat\right|_{\cal P}$ need not
coincide at
the intersections of different planes.

At this point lemma 2 comes to rescue. It is valid in any
dimension and may be applied to
$\left.U\right|_{\cal P}$ and
$\left.\bfieldlat\right|_{\cal P}$ (in place of $U$ and
$\bfieldlat$). We then conclude that
$$
  \left[\left.U\right|_{\cal P}\right]^{\temptrans}=
  \left.\bfieldlat\right|_{\cal P},
  \eqno\enum
$$
where $\temptrans$ is the
transformation defined by eq.(B.3).
Since $\temptrans(x)$ is independent of
the plane $\cal P$ passing through $x$, we have thus
shown that
$U^{\temptrans}=\bfieldlat$.
\endproof

\subsection B.3 Group theoretical lemmas

To be able to work out the consequences
of the lattice field equations later on, some technical
preparation is needed.

The eigenvalues $\lambda_{\alpha}$ of
any matrix $u\in\SU$ lie on the unit circle and their product is
equal to 1. We say that $u$ belongs to
the set $\setS$, if
$$
  \lambda_{\alpha}=\rme^{i\chi_{\alpha}},
  \quad\hbox{where}\quad\left|\chi_{\alpha}\right|<\pi/2
  \quad\hbox{and}\quad\sum_{\alpha=1}^N\chi_{\alpha}=0.
  \eqno\enum
$$
It is obvious that
$\setS$ is an open neighborhood of the group identity.
A useful criterion for a matrix $u$ to be an element of $\setS$
is

\proclaim
Lemma 4. Any matrix $u\in\SU$ with
$\Re\tr(1-u)<\min\left\{1,16/N\right\}$ is contained
in $\setS$.

\proof
The eigenvalues $\lambda_{\alpha}$ of $u$ may always be
parametrized by a set of angles
$\chi_{\alpha}$ such that
$$
  \lambda_{\alpha}=\rme^{i\chi_{\alpha}},
  \qquad-\pi<\chi_{\alpha}\leq\pi,
  \qquad\sum_{\alpha=1}^N\chi_{\alpha}=2\pi n,
  \eqno\enum
$$
where $n$ is an integer.
{}From the bound
$$
  \Re\tr(1-u)=\sum_{\alpha=1}^N(1-\cos\chi_{\alpha})<1
  \eqno\enum
$$
we conclude that
$\left|\chi_{\alpha}\right|<\pi/2$.
To prove that the sum of the angles $\chi_{\alpha}$ vanishes,
we first note that
$$
  1-\cos\chi\geq(2\chi/\pi)^2\quad\hbox{if}\quad
  \left|\chi\right|<\pi/2.
  \eqno\enum
$$
The sequence of inequalities
$$
  \left|\sum_{\alpha=1}^N\chi_{\alpha}\right|^2\leq
  N\sum_{\alpha=1}^N\chi_{\alpha}^2\leq
  {N\pi^2\over4}\Re\tr(1-u)<4\pi^2
  \eqno\enum
$$
is then easily established. It follows from these that
$|n|<1$, and since $n$ must be an integer, we are left
with $n=0$ as the only possible value.
\endproof

The significance of the set $\setS$ is elucidated by

\proclaim
Lemma 5. Any two matrices
$u,v\in\setS$ with the same traceless
anti-hermitean parts, viz.
$$
  u-u^{\dagger}-{1\over N}\tr\left(u-u^{\dagger}\right)=
  v-v^{\dagger}-{1\over N}\tr\left(v-v^{\dagger}\right),
  \eqno\enum
$$
are equal.

\proof
Let $\lambda_{\alpha}=\rme^{i\chi_{\alpha}}$ be the
eigenvalues of $u$ as in eq.(B.11). The associated eigenvectors
are also
eigenvectors of $u-u^{\dagger}$, with eigenvalues
$2i\sin\chi_{\alpha}$.
Since
$\chi_{\alpha}\neq\chi_{\beta}$ implies
$\sin\chi_{\alpha}\neq\sin\chi_{\beta}$,
the eigenspaces of $u$ and $u-u^{\dagger}$ must
in fact coincide.
The same statement applies to the other pair
of matrices, $v$
and $v-v^{\dagger}$.
So if we take eq.(B.16) into account,
it follows that all four matrices can be simultaneously diagonalized.

Let us now choose a basis of simultaneous eigenvectors.
The eigenvalues of $v$ may be parametrized
in the same way as those of $u$
through a set of angles
$\psi_{\alpha}$.
Eq.(B.16) then reduces to
$$
  \sin\chi_{\alpha}-\sin\psi_{\alpha}=t,
  \eqno\enum
$$
where $t$ is independent of $\alpha$.
In particular,
if $t>0$ one infers that
$\chi_{\alpha}>\psi_{\alpha}$ for all $\alpha$
(recall that all angles are between
$-\pi/2$ and $\pi/2$). This is impossible, however,
because both sets of angles must sum up to zero.
The same argument excludes $t<0$ and so we are left
with $\chi_{\alpha}=\psi_{\alpha}$ as the only
acceptable solution of eq.(B.17).
The matrices $u$ and $v$ are hence equal.
\endproof

\subsection B.4 Proof of theorem 1 for $D=2$

{}From now on the discussion is restricted to the two-dimensional
theory.
To simplify the notation we set
$\angle_{1\alpha}=\angle_{\alpha}$ and
$\angleprime_{1\alpha}=\angleprime_{\alpha}$.
Since the set of all lattice gauge fields is a compact manifold,
the infimum of the action is attained by at least one
configuration $U$.
In the following we assume that $U$ is
such a field.
We then need to show that $U$
is gauge equivalent to $\bfieldlat$.

\proclaim
Lemma 6.
For all $x$ the plaquette matrix
$$
  \plaq(x)=U(x,0)U(x+a\hat{0},1)U(x+a\hat{1},0)^{-1}U(x,1)^{-1}
  \eqno\enum
$$
is contained in the set $\setS$.

\proof
Since $U$ is a minimal action configuration, we have
$$
  S[U]\leq S[\bfieldlat]\leq
  (g_0^2TL)^{-1}\sum_{\alpha=1}^N
  \left(\angleprime_{\alpha}-\angle_{\alpha}\right)^2.
  \eqno\enum
$$
If we remember that
$\anglevector$ and $\anglevectorprime$ are in the fundamental
domain, it is easy to prove that
the angle differences
$\psi_{\alpha}=\angleprime_{\alpha}-\angle_{\alpha}$
satisfy
$$
  \left|\psi_{\alpha}-\psi_{\beta}\right|<2\pi,
  \qquad
  \sum_{\alpha=1}^N\psi_{\alpha}=0.
  \eqno\enum
$$
Using these properties, the estimate
$$
  \sum_{\alpha=1}^N\psi_{\alpha}^2=
  {1\over2N}
  \sum_{\alpha,\beta=1}^N
  \left(\psi_{\alpha}-\psi_{\beta}\right)^2<2(N-1)\pi^2
  \eqno\enum
$$
may be derived.

The action $S[U]$ is a sum of non-negative contributions,
one from each (unoriented) plaquette.
Any one of these must be smaller than
the right hand side in eq.(B.19) and thus, taking eq.(B.21)
and the bound (5.9)
into account, one deduces that
$$
  \Re\tr\left(1-\plaq(x)\right)
  <\min\left\{1,16/N\right\}
  \eqno\enum
$$
for all $x$.
Lemma 4 now tells us that $\plaq(x)$ is contained in $\setS$.
\endproof

\proclaim
Lemma 7.
The plaquette field $\plaq(x)$ is covariantly constant,
i.e.~it satisfies
$$
  \plaq(x)=U(x,\mu)\plaq(x+a\hat{\mu})U(x,\mu)^{-1}
  \eqno\enum
$$
for all $x$ and directions $\mu$.

\proof
Being a configuration with least action, $U$ must be a solution of
the lattice field equations
(cf.~subsect.~4.3).
In two dimensions these are equivalent to the requirement that
the unitary matrices
$$
  u=\plaq(x)
  \quad\hbox{and}\quad
  v=U(x,\mu)\plaq(x+a\hat{\mu})U(x,\mu)^{-1}
  \eqno\enum
$$
have the same
traceless anti-hermitean parts, for all $x$ and directions
$\mu$. From the above we know that
$u$ and $v$ are contained in the set $\setS$ and lemma 5
thus implies that $u=v$, as was to be shown.
\endproof

The plaquette field
$\plaq(x)$ may be regarded as a particular
gauge transformation function.
Eq.(B.23) then simply says that $U$ is invariant under
this transformation. In particular,
the boundary field $\bvaluelat$ is left invariant
and so, by lemma 1, we conclude that
$\plaq(x)$ must be constant and diagonal at $x^0=0$.
In other words, we have
$$
  \left.\plaq(x)\right|_{x^0=0}=
  \pmatrix{\rme^{i\chi_1} & 0              & \ldots & 0              \cr
           0              & \rme^{i\chi_2} & \ldots & 0              \cr
           \vdots         & \vdots         & \ddots & \vdots         \cr
        0              & 0              & \ldots & \rme^{i\chi_N} \cr},
  \eqno\enum
$$
where the angles $\chi_{\alpha}$
may be chosen such that
$$
  \left|\chi_{\alpha}\right|<\pi/2
  \quad\hbox{and}\quad\sum_{\alpha=1}^N\chi_{\alpha}=0
  \eqno\enum
$$
(as a consequence of lemma 6).

Another important implication of eq.(B.23) is that
the eigenvalues of $\plaq(x)$
are the same for all $x$. In particular, the action
$$
  S[U]={2TL\over g_0^2a^4}
  \sum_{\alpha=1}^N
  \left(1-\cos\chi_{\alpha}\right)
  \eqno\enum
$$
is determined by the angles $\chi_{\alpha}$.

\proclaim
Lemma 8. There exists a permutation $\perm$ and
a set of
integers $n_{\alpha}$ such that
$$
  \chi_{\alpha}=
  {a^2\over TL}
  \left(\angleprime_{\perm(\alpha)}
  +2\pi n_{\alpha}-\angle_{\alpha}\right).
  \eqno\enum
$$
In the special case where
$\sigma$ is the identity and $n_{\alpha}=0$
for all $\alpha$, the configuration $U$ is gauge equivalent
to $\bfieldlat$.

\proof
We first pass to the temporal gauge by applying the
gauge transformation $\temptrans$ [eq.(B.3)].
The transformed field
$\widetilde{U}=U^{\temptrans}$
and the associated plaquette field $\widetilde{\plaq}$
satisfy
$$
  \eqalignno{
  \widetilde{U}(x,0)&=1
  \quad\hbox{for all $x$,}
  &\enum\cr
  \noalign{\vskip1ex}
  \widetilde{U}(x,1)&=\bvaluelat({\bf x},1),
  \qquad
  \widetilde{\plaq}(x)=\plaq(x)
  \quad\hbox{at $x^0=0$.}
  &\enum\cr}
$$
Since
$\widetilde{\plaq}(x)$ is covariantly constant, we conclude
that it must be equal to $\plaq(0)$ for all $x$.

Next we note that
$$
  \widetilde{\plaq}(x)=\widetilde{U}(x+a\hat{0},1)
  \widetilde{U}(x,1)^{-1},
  \eqno\enum
$$
which immediately leads to
$$
  \widetilde{U}(x,1)=
  \left[\plaq(0)\right]^{x^0/a}\,
  \bvaluelat({\bf x},1).
  \eqno\enum
$$
The matrices on the right hand side of
this equation are diagonal, i.e.~we have
shown that $\widetilde{U}$ is an Abelian solution
of the field equations similar to $\bfieldlat$.

Let us now consider the parallel transporter for a loop
winding
once around the world at $x^0=T$. In the temporal gauge
it may be computed directly using eq.(B.32), or we may
note that it is just the gauge transform of the parallel
transporter determined by
the original field $U$. We thus obtain the relation
$$
  \temptrans(x)\exp\left\{L\bvalue'_1\right\}\temptrans(x)^{-1}
  =\left[\plaq(0)\right]^{TL/a^2}
  \exp\left\{L\bvalue_1\right\}.
  \eqno\enum
$$
The eigenvalues of the matrix on the left
are the phase factors $\rme^{i\angleprime_{\alpha}}$.
Up to a possible reordering,
they must be equal to the diagonal elements of the matrix
on the other side of the equation
and so we conclude that
eq.(B.28) holds for some permutation $\perm$ and some
integers $n_{\alpha}$.
Furthermore, if $\perm$ happens to
be the identity and $n_{\alpha}=0$ for all $\alpha$,
it is immediately clear from
eq.(B.32) that $\widetilde{U}=\bfieldlat$, i.e.~$U$
and $\bfieldlat$ are gauge equivalent in this case.
\endproof

The upshot then is that the possible minimal action
configurations $U$ are labelled by a permutation $\perm$
and a set of integers $n_{\alpha}$. Not all configurations
$(\perm,n)$ can occur,
but only those for which the angles
(B.28) satisfy the constraints (B.26).
These will be called {\it admissible}\/ in the following.
In particular,
the {\it trivial}\/ configuration,
where $\sigma$ is the identity and $n_{\alpha}=0$ for all
$\alpha$, is admissible. Note that the set of
admissible configurations $(\perm,n)$ is finite.

Taking eq.(B.27) and lemma 8 into account, the proof
of theorem 1 is now completed by

\proclaim
Lemma 9. On the set of admissible configurations $(\perm,n)$,
the minimum of the function
$$
  s(\perm,n)=\sum_{\alpha=1}^N
  \left(1-\cos\chi_{\alpha}\right)
  \eqno\enum
$$
is not degenerate and the unique minimizing configuration is the
trivial one.

\proof
Let us assume that $(\perm,n)$ is a non-trivial admissible
configuration. We then show that there exists another
admissible configuration
$(\widetilde{\perm},\widetilde{n})$, such that
$s(\perm,n)>s(\widetilde{\perm},\widetilde{n})$.

We first consider a
configuration $(\perm,n)$ where not all
$n_{\alpha}$'s are equal to zero.
Since $(\perm,n)$ is admissible,
the integers $n_{\alpha}$ must add up to zero. It is, therefore,
possible to find two indices
$\alpha$ and $\beta$ such that
$$
  n_{\alpha}\geq1
  \quad\hbox{and}\quad
  n_{\beta}\leq-1.
  \eqno\enum
$$
Let us now define a new configuration
$(\widetilde{\perm},\widetilde{n})$ through
$$
  \widetilde{\perm}({\gamma})=\cases{
  \perm({\beta})     & if $\gamma=\alpha$,  \cr
  \noalign{\vskip0.5ex}
  \perm({\alpha})    & if $\gamma=\beta$,   \cr
  \noalign{\vskip0.5ex}
  \perm({\gamma})    & otherwise,           \cr}
  \eqno\enum
$$
and
$$
  \widetilde{n}_{\gamma}=\cases{
  n_{\beta }+1   & if $\gamma=\alpha$,  \cr
  \noalign{\vskip0.5ex}
  n_{\alpha}-1   & if $\gamma=\beta$,   \cr
  \noalign{\vskip0.5ex}
  n_{\gamma}     & otherwise.          \cr}
  \eqno\enum
$$
The associated angles
$\widetilde{\chi}_{\gamma}$ are equal to $\chi_{\gamma}$
with the exception of
$$
  \eqalignno{
  \widetilde{\chi}_{\alpha}&=
  {a^2\over TL}
  \left[\angleprime_{\perm(\beta )}+2\pi\left(n_{\beta }+1\right)
  -\angle_{\alpha}\right],
  &\enum\cr
  \noalign{\vskip1ex}
  \widetilde{\chi}_{\beta }&=
  {a^2\over TL}
  \left[\angleprime_{\perm(\alpha)}+2\pi\left(n_{\alpha}-1\right)
  -\angle_{\beta }\right].
  &\enum\cr}
$$
Note that $\widetilde{\perm}$ is just
a transposition of $\alpha$ and $\beta$ followed by $\perm$.

It is not difficult to show that
$(\widetilde{\perm},\widetilde{n})$
is an admissible configuration. Furthermore, a little algebra
yields
$$
  \eqalignno{
  &s(\perm,n)-s(\widetilde{\perm},\widetilde{n})=
  &\enum\cr
  \noalign{\vskip1ex}
  &4\cos\left[\frac{1}{4}\left(\chi_{\alpha}+\chi_{\beta}
  +\widetilde{\chi}_{\alpha}+\widetilde{\chi}_{\beta}\right)\right]
  \sin\left[\frac{1}{2}\left(\chi_{\alpha}
  -\widetilde{\chi}_{\alpha}\right)\right]
  \sin\left[\frac{1}{2}\left(\chi_{\alpha}
  -\widetilde{\chi}_{\beta }\right)\right].
  &\cr}
$$
The arguments of all
trigonometric functions in this expression
are less than $\pi/2$ in magnitude. In particular, the first factor
is positive.
Concerning the other two factors, we note that their arguments
$$
  \eqalignno{
  \chi_{\alpha}-\widetilde{\chi}_{\alpha}&=
  {a^2\over TL}
  \left[\angleprime_{\perm(\alpha)}-\angleprime_{\perm(\beta )}
  +2\pi\left(n_{\alpha}-n_{\beta }-1\right)\right],
  &\enum\cr
  \noalign{\vskip1ex}
  \chi_{\alpha}-\widetilde{\chi}_{\beta }&=
  {a^2\over TL}
  \left[2\pi+\angle_{\beta }-\angle_{\alpha}\right],
  &\enum\cr}
$$
are positive, because the angle vectors $\anglevector$ and
$\anglevectorprime$ are in the fundamental domain and because
the integers $n_{\alpha}$ and $n_{\beta}$ satisfy the bounds (B.35).
So we conclude that
the action of the new configuration
$(\widetilde{\perm},\widetilde{n})$ is strictly lower
than $s(\perm,n)$.

Let us now consider a configuration $(\perm,n)$, where
$\perm$ is non-trivial but where all $n_{\alpha}$'s vanish.
All configurations of this type are admissible.
Since $\perm$ is not the identity, there are two indices
$\alpha$ and $\beta$ such that
$$
  \alpha<\beta
  \quad\hbox{and}\quad
  \perm(\alpha)>\perm(\beta).
  \eqno\enum
$$
The new configuration
$(\widetilde{\perm},\widetilde{n})$ in the present case
is given by eq.(B.36) and $\widetilde{n}_{\gamma}=0$ for all
$\gamma$.
Eq.(B.40) is then still valid and in view of
$$
  \eqalignno{
  \chi_{\alpha}-\widetilde{\chi}_{\alpha}&=
  {a^2\over TL}
  \left[\angleprime_{\perm(\alpha)}-\angleprime_{\perm(\beta )}\right],
  &\enum\cr
  \noalign{\vskip1ex}
  \chi_{\alpha}-\widetilde{\chi}_{\beta }&=
  {a^2\over TL}
  \left[\angle_{\beta }-\angle_{\alpha}\right],
  &\enum\cr}
$$
all factors on the right hand side are positive.
It follows that
$s(\perm,n)>s(\widetilde{\perm},\widetilde{n})$
and we have thus proved the lemma.
\endproof


\appendix C

In this appendix we derive a useful expression
for the determinant of a general
second order difference operator
acting on a space of wave functions with Dirichlet boundary conditions.
This result is needed
in sect.~7 to compute the determinants
of $\deltazero$ and $\deltaone$ in a constant
Abelian background field.
The basic idea is borrowed
from Coleman's Erice lecture on the uses of instantons
where he treats the analogous case
of a second order differential operator
(see ref.[\ref{Coleman}], p.340).

\subsection C.1 Definitions

The general
second order difference operator $\Delta$ acts on complex
wave functions $\psi(t)$ with $n$ components,
defined at integer values of $t$.
Explicitly, $\Delta$ is given by
$$
  \Delta\psi(t)=
  \amat(t)\psi(t+1)+\bmat(t)\psi(t)+\cmat(t)\psi(t-1),
  \eqno\enum
$$
where $\amat$, $\bmat$ and $\cmat$ are some complex
$n\times n$ matrices depending on $t$.

The wave functions which are
defined for $t=0,1,2,\ldots,T$ and satisfy
Dirichlet boundary conditions,
$$
  \psi(t)=0
  \quad\hbox{at $t=0$ and $t=T$,}
  \eqno\enum
$$
form a vector space $\funcspace$ of dimension
$\dim=n(T-1)$. If $\psi(t)$ is an element of this space,
eq.(C.1) is meaningful for $0<t<T$ and $\Delta$
may thus be regarded as
a linear operator in $\funcspace$.

In the following
we assume that $\amat(t)$ is invertible
for all $t$ and that $\Delta$ is hermitean relative to
some scalar product on $\funcspace$.
In particular, we take it for granted that
there exists a complete set of eigenfunctions.

\subsection C.2 Statement of result

For any complex $\lambda$ the equation
$$
  (\Delta-\lambda)\psi(t)=0,
  \qquad t>0,
  \eqno\enum
$$
has a unique solution
with $\psi(0)=0$ and prescribed initial value at $t=1$.
It can be computed recursively
by first solving eq.(C.3)
at $t=1$, then at $t=2$, and so on.
After $T-1$ steps
one obtains $\psi(T)$ which
is in general not equal to zero.
It is obvious, however, that $\psi(T)$ depends linearly
on the initial value $\psi(1)$,
i.e.~there exists an $n\times n$ matrix
$\solmat(\lambda)$ such that
$$
  \psi(T)=\solmat(\lambda)\psi(1).
  \eqno\enum
$$
This matrix is evidently determined through the
coefficients $\amat$, $\bmat$ and $\cmat$.
It is comparatively easy to evaluate, requiring
a computational effort proportional to $n^3T$.

The formula alluded to above now reads
$$
  \det\left(\Delta-\lambda\right)=
  \det\solmat(\lambda)\prod_{t=1}^{T-1}\det\left[-\amat(t)\right],
  \eqno\enum
$$
where $\Delta$ is here considered to be
an operator in $\funcspace$, as described above.
In particular, the determinant of $\Delta$ is obtained
if we set $\lambda=0$.

\subsection C.3 Proof of eq.(C.5)

It is easy to show
that the matrix $\solmat(\lambda)$ is a polynomial in
$\lambda$ of degree $T-1$.
The leading term is
$$
  \solmat(\lambda)=\lambda^{T-1}
  \left\{\amat(1)\amat(2)\ldots\amat(T-1)\right\}^{-1}
  +\rmO(\lambda^{T-2}).
  \eqno\enum
$$
It follows from these remarks that
$$
  P(\lambda)=
  \det\solmat(\lambda)\prod_{t=1}^{T-1}\det\left[-\amat(t)\right]
  \eqno\enum
$$
is a polynomial in $\lambda$ too, with leading term equal to
$(-\lambda)^{\dim}$.

Let us now assume
that $\mu$ is an eigenvalue of $\Delta$ with multiplicity
$k$. If $\psi\in\funcspace$
is one of the associated eigenfunctions, we have
$$
  0=\psi(T)=\solmat(\mu)\psi(1).
  \eqno\enum
$$
The matrix $\solmat(\mu)$ thus has
a zero mode and we conclude
that $P(\mu)=0$.
Actually, since there
are $k$ linearly independent eigenfunctions, we may choose
a basis such that
the first $k$ columns of $\solmat(\mu)$ vanish.
The multiplicity of the zero of
$P(\lambda)$ at $\lambda=\mu$
is, therefore, greater or equal to $k$.

The total number of eigenvalues of $\Delta$, including multiplicities,
is equal to $\dim$. Since this is also the degree of
$P(\lambda)$, it follows that
this polynomial cannot have any further zeros
besides the eigenvalues of $\Delta$ and that, moreover,
the corresponding multiplicities must coincide.

We have thus shown that $P(\lambda)$ is
the characteristic polynomial of the operator $\Delta$,
which is precisely the content of eq.(C.5).


\appendix D

In this appendix the matrices
$\amat_{\mu\nu}$, $\bmat_{\mu\nu}$ and $\cmat_{\mu\nu}$
introduced in subsect.~7.4 are given explicitly.
They are all real and satisfy
$$
  \bmat_{\mu\nu}(t)=\bmat_{\nu\mu}(t),
  \qquad
  \cmat_{\mu\nu}(t)=\amat_{\nu\mu}(t-1),
  \eqno\enum
$$
as required for a symmetric operator.
In the following we shall use the abbreviations
$$
  \eqalignno{
  \phat_k&=
  2\sin\left[\frac{1}{2}p_k\right],
  &\enum\cr
  \noalign{\vskip1.5ex}
  s_k(t)&=
  2\sin\left[\frac{1}{2}\left(p_k+\beta(t)\right)\right].
  &\enum\cr}
$$
We now list the independent elements of the matrices
which describe the action of $\deltaone$ in
the $\fsp{0}$ sector.
$$
  \eqalignno{
  \amat_{00}(t)&=-\lambda_0,
  &\enum\cr
  \noalign{\vskip1ex}
  \amat_{kl}(t)&=-\cgh\,\delta_{kl},
  &\enum\cr
  \noalign{\vskip1ex}
  \amat_{0k}(t)&=-\left[\cgh-\lambda_0\right]\phat_k,
  &\enum\cr
  \noalign{\vskip1ex}
  \amat_{k0}(t)&=0,
  &\enum\cr
  \noalign{\vskip2ex}
  \bmat_{00}(t)&=2\lambda_0+\cgh\,\phat_j\phat_j,
  &\enum\cr
  \noalign{\vskip1ex}
  \bmat_{kl}(t)&=\left[2\cgh+\phat_j\phat_j\right]\delta_{kl}
  +(\lambda_0-1)\phat_k\phat_l,
  &\enum\cr
  \noalign{\vskip1ex}
  \bmat_{0k}(t)&=
  \left[\cgh-\lambda_0\right]\phat_k,
  &\enum\cr}
$$
The matrices describing the action
of $\deltaone$ in the $\fsp{+}$ subspace
are as follows.
$$
  \eqalignno{
  \amat_{00}(t)&=-\lambda_0,
  &\enum\cr
  \noalign{\vskip1ex}
  \amat_{kl}(t)&=-\delta_{kl},
  &\enum\cr
  \noalign{\vskip1ex}
  \amat_{0k}(t)&=\lambda_0s_k(t+1)-s_k(t),
  &\enum\cr
  \noalign{\vskip1ex}
  \amat_{k0}(t)&=0,
  &\enum\cr
  \noalign{\vskip2ex}
  \bmat_{00}(t)&=2\lambda_0+s_j(t)s_j(t+1),
  &\enum\cr
  \noalign{\vskip1ex}
  \bmat_{kl}(t)&=\left[2\cgh+s_j(t)s_j(t)\right]\delta_{kl}
  +(\lambda_0-1)s_k(t)s_l(t),
  &\enum\cr
  \noalign{\vskip1ex}
  \bmat_{0k}(t)&=
  s_k(t+1)-\lambda_0s_k(t).
  &\enum\cr}
$$


\beginbibliography

\bibitem{SymSchrodinger}
K. Symanzik,
Nucl. Phys. B190 [FS3] (1981) 1

\bibitem{LueSchrodinger}
M. L\"uscher,
Nucl. Phys. B254 (1985) 52

\bibitem{LueWeWo}
M. L\"uscher, P. Weisz and U. Wolff,
Nucl. Phys. B359 (1991) 221

\bibitem{DeWittBackground}
B. S. DeWitt,
Phys. Rev. 162 (1967) 1195 and 1239

\bibitem{Honerkamp}
J. Honerkamp,
Nucl. Phys. B48 (1972) 269

\bibitem{HooftBackgroundI}
G. 't Hooft,
Nucl. Phys. B62 (1973) 444

\bibitem{HooftBackgroundII}
G. 't Hooft,
The Background Field Method in Gauge Field Theory,
Lecture given at Karpacz (1975),
{\sl in}\/: Acta Universitatis Wratislaviensis,
No. 368 (1976) 345

\bibitem{Abbott}
L. F. Abbott,
Nucl. Phys. B185 (1981) 189

\bibitem{DashenGross}
R. Dashen and D.J. Gross,
Phys. Rev. D23 (1981) 2340

\bibitem{KollerBaal}
J. Koller and P. van Baal,
Nucl. Phys. B302 (1988) 1

\bibitem{BaalBackground}
P. van Baal,
Phys. Lett. B224 (1989) 397;
Nucl. Phys. B (Proc. Suppl.) 17 (1990) 581;
Nucl. Phys. B351 (1991) 183

\bibitem{VenBackground}
A. E. M. van de Ven,
Two Loop Quantum Gravity,
DESY 91-115 (1991)

\bibitem{WoGbar}
U. Wolff,
Nucl. Phys. B265 [FS15] (1986) 506 and 537

\bibitem{BePo}
A. A. Belavin, A. M. Polyakov, A. S. Schwartz and
Yu. S. Tyupkin,
Phys. Lett. 59B (1975) 85

\bibitem{JaRe}
R. Jackiw and C. Rebbi,
Phys. Rev. Lett. 37 (1976) 172

\bibitem{CaDaGr}
C. G. Callan, R. F. Dashen and D. J. Gross,
Phys. Lett. 63B (1976) 334

\bibitem{tHooftInstA}
G. 't Hooft,
Phys. Rev. Lett. 37 (1976) 8

\bibitem{tHooftInstB}
G. 't Hooft,
Phys. Rev. D14 (1976) 3432 [E: Phys. Rev. D18 (1978) 2199]

\bibitem{BelavinInst}
A. A. Belavin and A. M. Polyakov,
Nucl. Phys. B123 (1977) 429

\bibitem{SchwarzInst}
A. S. Schwarz,
Commun. Math. Phys. 64 (1979) 233

\bibitem{OsbornInst}
H. Osborn,
Ann. Phys. (N.Y.) 135 (1981) 373

\bibitem{MorrisInst}
T. R. Morris, D. A. Ross and C. T. Sachrajda,
Nucl. Phys. B255 (1985) 115

\bibitem{tHooftVeltman}
G. 't Hooft and M. Veltman,
Nucl. Phys. B44 (1972) 189

\bibitem{BoGia}
C. G. Bollini and J. J. Giambiagi,
Nuovo Cim. B12 (1972) 20

\bibitem{Ash}
J. F. Ashmore,
Nuovo Cim. Lett. 4 (1972) 289

\bibitem{CiMo}
G. M. Cicuta and E. Montaldi,
Nuovo Cim. Lett. 4 (1972) 329

\bibitem{LueDimReg}
M. L\"uscher,
Ann. Phys. (N.Y.) 142 (1982) 359

\bibitem{tHooftMS}
G. 't Hooft,
Nucl. Phys. B61 (1973) 455

\bibitem{Jones}
D. R. T. Jones,
Nucl. Phys. B75 (1974) 531

\bibitem{Caswell}
W. E. Caswell,
Phys. Rev. Lett. 33 (1974) 244

\bibitem{Tarasov}
O. V. Tarasov, A. A. Vladimirov and A. Yu. Zharkov,
Phys. Lett. 93B (1980) 429

\bibitem{DeWitt}
B. S. DeWitt,
Dynamical Theory of Groups and Fields
(Gordon and Breach, London, 1965)

\bibitem{McSi}
H. P. McKean and I. M. Singer,
J. Diff. Geom. 1 (1967) 43

\bibitem{Seeley}
R. T. Seeley,
Am. J. Math. 91 (1969) 889 and 963

\bibitem{BaBl}
R. Balian and C. Bloch,
Ann. Phys. (N.Y.) 60 (1970) 401; ibid 64 (1971) 271

\bibitem{Gil}
P. B. Gilkey,
Invariance Theory, the Heat Equation
and the Atiyah-Singer Index Theorem
(Publish or Perish, Wilmington, 1984)

\bibitem{LueSyWe}
M. L\"uscher, K. Symanzik and P. Weisz,
Nucl. Phys. B17396 (1980) 365

\bibitem{DuNiOlPe}
B. Durhuus, H. B. Nielsen, P. Olesen and J. L. Petersen,
Nucl. Phys. B196 (1982) 498

\bibitem{DuOlPe}
B. Durhuus, P. Olesen and J. L. Petersen,
Nucl. Phys. B198 (1982) 157

\bibitem{CoVaZe}
G. Cognola, L. Vanzo and S. Zerbini,
Phys. Lett. B241 (1990) 381

\bibitem{AvOs}
D. M. McAvity and H. Osborn,
Class. Quant. Grav. 8 (1991) 603

\bibitem{Wilson}
K. G. Wilson,
Phys. Rev. D10 (1974) 2445

\bibitem{WilsonKogut}
K. G. Wilson and J. B. Kogut,
Phys. Repts. 12 (1974) 75

\bibitem{WilsonTrans}
K. G. Wilson,
Quantum Chromodynamics on a Lattice,
Lectures given at Carg\`ese (1976),
{\sl in}\/:
New Developments in Quantum Field Theory and Statistical Mechanics,
ed. M. L\'evy and P. Mitter (Plenum, New York, 1977)

\bibitem{LueTrans}
M. L\"uscher,
Commun. Math. Phys. 54 (1977) 238

\bibitem{CreutzTrans}
M. Creutz,
Phys. Rev. D15 (1977) 1128; ibid D35 (1987) 1460

\bibitem{LueLesHouches}
M. L\"uscher,
Selected Topics in Lattice Field Theory,
Lectures given at Les Houches (1988),
{\sl in}\/: Fields, Strings and Critical Phenomena,
ed. E. Br\'ezin and J. Zinn-Justin
(North Holland, Amsterdam, 1989)

\bibitem{SymanzikLELa}
K. Symanzik,
Cutoff Dependence in Lattice $\phi^4_4$ Theory,
Lecture given at Carg\`ese (1979),
{\sl in}\/: Recent Developments in Gauge Theories,
ed. G. 't Hooft et al.
(Plenum, New York, 1980)

\bibitem{SymanzikLELb}
K. Symanzik,
Concerning the Continuum Limit in some Lattice Theories,
{\sl in}\/: 21st International Conference on High Energy Physics,
Paris (1982),
ed. P. Petiau and M. Porneuf,
J. de Physique 43 (1982) C3; 254

\bibitem{SymanzikImpA}
K. Symanzik,
Some Topics in Quantum Field Theory,
{\sl in}\/: Mathematical Problems in Theoretical Physics,
ed. R. Schrader et al., Lecture Notes in Physics,
Vol. 153 (Springer, Berlin, 1982)

\bibitem{SymanzikImpB}
K. Symanzik,
Nucl. Phys. B226 (1983) 187 and 205

\bibitem{Keller}
G. Keller,
The Perturbative Construction of Symanzik's Improved
Action for $\phi^4_4$ and ${\rm QED}_4$,
MPI-PTh/92-6 (1992)

\bibitem{WeiszImp}
P. Weisz,
Nucl. Phys. B212 (1983) 1

\bibitem{WeiszWohlert}
P. Weisz and R. Wohlert,
Nucl. Phys. B236 (1984) 397

\bibitem{CuMePa}
G. Curci, P. Menotti and G. Paffuti,
Phys. Lett. 130B (1983) 205

\bibitem{LueWeImpA}
M. L\"uscher and P. Weisz,
Commun. Math. Phys. 97 (1985) 59
[E: Commun. Math. Phys. 98 (1985) 433]

\bibitem{LueImp}
M. L\"uscher, Improved Lattice Gauge Theories,
Seminar given at Les Houches (1984),
{\sl in}\/: Critical Phenomena, Random Systems, Gauge Theories,
ed. K. Osterwalder and R. Stora (North Holland, Amsterdam, 1986)

\bibitem{LueWeImpB}
M. L\"uscher and P. Weisz,
Phys. Lett. 158B (1985) 250; Nucl. Phys. B266 (1986) 309

\bibitem{GJK}
A. Gonzalez-Arroyo, J. Jurkiewicz and C. P. Korthals-Altes,
Ground State Metamorphosis for Yang-Mills Fields on a Finite Periodic
Lattice,
Lecture given at Freiburg (1981),
{\sl in}\/: Structural Elements in Particle Physics and
Statistical Mechanics, ed. J. Honerkamp and K. Pohlmeyer
(Plenum, New York, 1983)

\bibitem{LueTorons}
M. L\"uscher,
Nucl. Phys. B219 (1983) 233

\bibitem{Coleman}
S. Coleman,
Aspects of Symmetry
(Cambridge University Press, Cambridge, 1985)

\bibitem{Hasenfratz}
A. Hasenfratz and P. Hasenfratz,
Phys. Lett. 93B (1980) 165; Nucl. Phys. B193 (1981) 210

\bibitem{Weisz}
P. Weisz,
Phys. Lett. 100B (1981) 331

\endbibliography

\bye